\documentclass[modern]{aastex631}

\usepackage{graphicx}
\usepackage{booktabs}
\def\gaia{\textsl{Gaia}}

\usepackage[T1]{fontenc}
\usepackage[utf8]{inputenc}
\usepackage{CJKutf8}
\usepackage{xcolor}
\usepackage{listings}
\usepackage{amsmath}
\usepackage{comment}
\usepackage{tabularx}
\usepackage{booktabs}
\definecolor{mauve}{rgb}{0.88, 0.69, 1.0}
\definecolor{oldmauve}{rgb}{0.4, 0.19, 0.28}
\newcommand{\msun}{M$_{\odot}$}              
\newcommand{\singlechi}{$\chi^2_{\mathrm{single}}$}
\newcommand{\binarychi}{$\chi^2_{\mathrm{WDMS}}$}

\begin{document}

\title{Identification of 30,000 White Dwarf-Main Sequence binaries candidates from Gaia DR3 BP/RP(XP) low-resolution spectra}

\author[0000-0002-3651-5482]{Jiadong Li}
\affiliation{Max-Planck-Institut für Astronomie, Königstuhl 17, D-69117 Heidelberg, Germany}

\author[0000-0001-5082-9536]{Yuan-Sen Ting}
\affiliation{Department of Astronomy, The Ohio State University, Columbus, OH 43210, USA}
\affiliation{Center for Cosmology and AstroParticle Physics (CCAPP), The Ohio State University, Columbus, OH 43210, USA}
\affiliation{Max-Planck-Institut für Astronomie, Königstuhl 17, D-69117 Heidelberg, Germany}

\author[0000-0003-4996-9069]{Hans-Walter Rix}
\affiliation{Max-Planck-Institut für Astronomie, Königstuhl 17, D-69117 Heidelberg, Germany}

\author[0000-0001-5417-2260]{Gregory M. Green}
\affiliation{Max-Planck-Institut für Astronomie, Königstuhl 17, D-69117 Heidelberg, Germany}

\author[0000-0003-2866-9403]{David W. Hogg}
\affiliation{Center for Cosmology and Particle Physics, Department of Physics, New York University, USA}
\affiliation{Max-Planck-Institut für Astronomie, Königstuhl 17, D-69117 Heidelberg, Germany}
\affiliation{Flatiron Institute, New York, USA}

\author[0000-0003-3243-464X]{Juan-Juan Ren}
\affiliation{Key Laboratory of Space Astronomy and Technology, National Astronomical Observatories, Chinese Academy of Sciences, Beijing 100012, China}

\author[0000-0001-9590-3170]{Johanna M\"uller-Horn}
\affiliation{Max-Planck-Institut für Astronomie, Königstuhl 17, D-69117 Heidelberg, Germany}

\author[0000-0001-8898-9463]{Rhys Seeburger}
\affiliation{Max-Planck-Institut für Astronomie, Königstuhl 17, D-69117 Heidelberg, Germany}

\begin{abstract}
White dwarf-main sequence (WDMS) binary systems are essential probes for understanding binary stellar evolution and play a pivotal role in constraining theoretical models of various transient phenomena. 
In this study, we construct a catalog of WDMS binaries using Gaia DR3's low-resolution BP/RP (XP) spectra.
Our approach integrates a model-independent neural network for spectral modelling with Gaussian Process Classification to accurately identify WDMS binaries among over 10 million stars within 1 kpc. 
This study identify approximately 30,000 WDMS binary candidates, including 1,700 high-confidence systems confirmed through spectral fitting. 
Our technique is shown to be effective at detecting systems where the main-sequence star dominates the spectrum - cases that have historically challenged conventional methods.
Validation using GALEX photometry reinforces the reliability of our classifications: 70\% of candidates with an absolute magnitude $M_{G} > 7$ exhibit UV excess, a characteristic signature of white dwarf companions.
Our all-sky catalog of WDMS binaries expands the available dataset for studying binary evolution and white dwarf physics and sheds light on the formation of WDMS.
\end{abstract}

\section{Introduction} \label{sec:intro}

Binary star systems containing a white dwarf (WD) and a main sequence (MS) companion (WDMS binaries) are fundamental laboratories for studying stellar evolution in interacting systems. 
The evolution of WDMS binaries follows two distinct pathways, determined by their initial orbital separations \citep{deKool1993, Willems2004}. 
In approximately 75\% of the cases, the initial separations are sufficiently wide ($\gtrsim$ 10 AU) that the more massive star evolves essentially as a single star, eventually becoming a white dwarf without interaction with the companion \citep{Farihi2010}.
These systems maintain orbital separations similar to their initial configurations.
The remaining $\sim$25\% of systems experience a more dramatic evolution. 
As the more massive star expands into the giant phase, the binary undergoes a Common Envelope (CE) phase in which orbital energy and angular momentum losses lead to a dramatic decrease in orbital separation \citep{Iben1993}, although recently putative PCEBs have been found in wide orbits \citep{Yamaguchi2024}.
This process results in the ejection of the envelope and the formation of a post-common envelope binary (PCEB). 
The PCEBs play an important role in a wide range of problems, e.g., constraining the effeciency of common evelope evolution \citep{Rebassa2012, Zorotovic2014},
These PCEBs are the progenitors of various exotic systems, including Type Ia supernovae (e.g., \citealt{Han2004,Wang2012, Hernandez2022}) , double degenerate binaries, cataclysmic variables \citep{Parsons2013, Sun2021}, and low-mass X-ray binaries \citep{Zorotovic2010}. 

The identification and characterization of WDMS binaries have historically been challenging due to the vast difference in luminosities between components,  particularly in systems with AFGK-type stars.
White dwarfs are typically much fainter than their main sequence companions, making their spectral signatures difficult to detect in unresolved systems. 
Different techniques have been developed to identify these systems: spatially resolved wide binaries can be detected through common-proper motion pairs, while unresolved systems require spectroscopic analysis or the detection of eclipses. 
Using common-proper motion, \gaia\ \citep{GaiaCollaboration2016} EDR3 identified approximately 16,000 wide WDMS systems within 1 kpc \citep{El-Badry2021}. 
For unresolved systems where one of the components dominates the spectral energy distribution (estimated to be ~90\% of cases), identification becomes challenging \citep{Rebassa2021}. 

Large spectroscopic surveys have made great advances in identifying WDMS binaries, with the Sloan Digital Sky Survey (SDSS; \citealt{York2000}) identifying over 3,200 WDMS systems \citep{Rebassa2016} and the Large Sky Area Multi-Object Fiber Spectroscopic Telescope (LAMOST; \citealt{Deng2012}) survey discovering approximately 1,000 additional pairs \citep{Ren2014, Ren2018} by spectra with resolution $\sim 2000$.
These samples have been explored to study CE evolution \citep{Zorotovic2010} as well as to test mass-radius relationships \citep{Parsons2017}.

However, both SDSS and LAMOST samples are subject to selection effects and are particularly biased against systems where one component dominates the spectral energy distribution (SED) \citep{Rebassa2010}.  
These limitations manifest themselves in two key aspects. 
First, the traditional identification methods, which rely on $\chi^2$-fitting, wavelet transforms, and human visual inspection, struggle to identify the WDMS that has a component dominated the SED. 
Recent studies have shown that such systems can comprise up to ~91\% of the total population of WDMS binaries \citep{Rebassa2021}, making this a limitation in our understanding of these systems.
Recent work combining UV and optical data has helped identify some previously missed hot WD systems \citep{Nayak2024}, many WDMS binaries likely remain undetected.
Second, existing established approaches to identifying WDMS binaries have their own challenges. They either rely on previously identified SDSS/LAMOST systems as training sets, potentially inheriting the same selection biases \citep{Rebassa2021}, or depend on synthetic spectral libraries to construct theoretical WDMS binary templates \citep{Echeverry2022}. 
Although theoretical models have made progress for cool dwarf stars \citep{Allard2011}, there remain some systematic differences between synthetic and observed spectra, especially for cool dwarf stars \citep{jdli2021, Qiu2024}, which can affect the accuracy of binary identification.
These model limitations can lead to erroneous identification and characterization. 

The \gaia\;mission \citep{GaiaCollaboration2016} opens a new window in the study of WDMS binaries \citep{Nayak2024, Grondin2024} over a wide range of stellar parameters and distances \citep{Echeverry2022}, with well-understood selection effects and homogeneous data quality.

In this paper, we present a new method for detecting WDMS binaries using \gaia\ DR3 XP spectra. 
Our approach combines the interpretability of $\chi^2$ fitting with machine learning techniques. 
We first develop data-driven spectral emulators for single main-sequence stars, white dwarfs, and WDMS binaries, following a similar iterative approach to eliminate binary contamination in the training sample as described by \citet{El-Badry2018}.
This ensures that our models are built from observations themselves, independent of theoretical spectral templates. 
We then employ a Gaussian Process Classifier (GPC) trained on these candidates to identify WDMS binaries across the full parameter space - particularly targeting systems where one component dominates the spectral energy distribution, which are challenging to detect through $\chi^2$ fitting alone.
Combining the high-quality homogeneous data provided by \gaia\ with this hybrid methodology, we aim to construct the most complete sample of WDMS binaries from spectrophotometric data from \gaia.

The remainder of the paper is structured as follows:
In Section \ref{sec:data}, we describe the \gaia\ DR3 data and our sample selection criteria. 
Section \ref{sec:method} outlines our spectral fitting methodology and the development of our single-star and binary models. 
In Section \ref{sec:gpc}, we present our classification approach using our Gaussian Process classification techniques. 
Section \ref{sec:discussion} details the results of our WDMS binary detection, including a discussion of the reliability of our method. 
We summarize our findings and discuss their implications for future studies of WDMS systems and binary star populations.

\section{Data}\label{sec:data}

In this section, we first describe the properties and processing of the \textit{Gaia} BP/RP spectra, and then we detail our sample selection criteria and pre-processing steps. 
Furthermore, we present a detailed account of the handling of extinction corrections and the creation of clean training samples for both white dwarfs and main-sequence stars.

\subsection{XP: the Gaia BP/RP low-resolution spectra}\label{subsec:xp}
The third data release of the \gaia\ mission (DR3) \citep{GaiaCollaboration2022} has delivered spectrophotometric data through its low-resolution measurements. 
The spectra are obtained using two dedicated instruments: the ``Blue Photometer'' (BP, 330-680 nm) and ``Red Photometer'' (RP, 640-1050 nm). 
Together, these instruments -- hereafter referred to as BP/RP (XP) -- have collected a dataset comprising 78 billion individual transit observations. 
The \gaia\ data processing pipeline calibrates the resulting 65 billion epoch spectra \citep{DeAngeli2022}, producing high-quality spectrophotometric data for approximately 220 million stars in DR3 \citep{Montegriffo2022}.

Unlike classical spectra that provide discrete flux measurements at specific wavelengths, XP spectra are expressed as continuous functions through a basis function approach \citep{DeAngeli2022}. 
The BP and RP spectral data are projected onto a set of 55 orthonormal Hermite functions, yielding a compact 110-dimensional coefficient vector representation of the complete spectrum.
This encoding method employs a hierarchical coefficient structure: the primary coefficients encode the bulk of the spectral flux distribution, while higher-order coefficients capture detailed spectral features \citep{DeAngeli2022}. 
Through this mathematical framework, the XP spectra achieve data compression while preserving essential spectroscopic information. 
We convert the coefficients back to wavelength space for our analysis (as discussed in subsection \ref{subsec:sample_xp}) since spectral features are more readily interpreted in wavelength units.

\subsection{Data Selection}\label{subsec:data_selection}

Given the magnitude limitations of \gaia\, XP, we restrict our analysis to stars within 1 kpc. 
To construct our initial sample, we execute the following query on the \gaia\ DR3 database:

\begin{lstlisting}
SELECT *
FROM gaiadr3.gaia_source
WHERE parallax > 1
  AND phot_bp_mean_mag < 18
  AND bp_rp BETWEEN -0.5 AND 5
  AND parallax_over_error > 10
  AND phot_g_mean_mag + 5*log10(abs(parallax)/100) > 6
  AND ruwe < 1.4
  AND has_xp_continuous = `true'
\end{lstlisting}
Our data process consists of the following steps:

First, we selected every stellar object that have XP spectra within approximately 1 kpc (parallax $>$ 1 mas), with a G-band magnitude brighter than 20.0 mag which has BP/RP (XP) spectra, and a B-R color index between -0.5 and 5. 
Second, we addressed a systematic bias in the \gaia\ photometry affecting faint red sources, where the mean $G_{\rm{BP}}$ (hereafter, B) flux is known to be overestimated. 
This bias arises from the data processing pipeline, which excludes epochs with calibrated fluxes below $1~e^{-}s^{-1}$ when computing the weighted mean flux \citep{rielloGaiaEarlyData2021}. 
The exclusion occurs because measurements below this threshold deviate from a normal distribution, potentially skewing the flux estimates. 
To ensure photometric reliability, we imposed a magnitude limit of B$<18$ mag, where this systematic effect remains negligible and flux measurements maintain their statistical integrity.
Third, we selected samples with un-derreddened absolute magnitude $M_G$ fainter than 6 to focus on low-mass stars.

We applied magnitude constraints based on parallax and retained only sources with a renormalized unit weight error (RUWE) $<1.4$ and a parallax-over-error ratio $>10$ to ensure robust astrometric measurements. 
After these selections, we obtain 10,163,728 stars with unique \gaia\ DR3 source identifiers. 
While this initial RUWE cut was implemented to build clean single-star templates for MS and WD stars, when we later applied our method to the full dataset without the RUWE restriction, we added an additional 669,380 stars with RUWE $>1.4$.

\begin{figure*}
    \centering
    \includegraphics[width=0.32\linewidth]{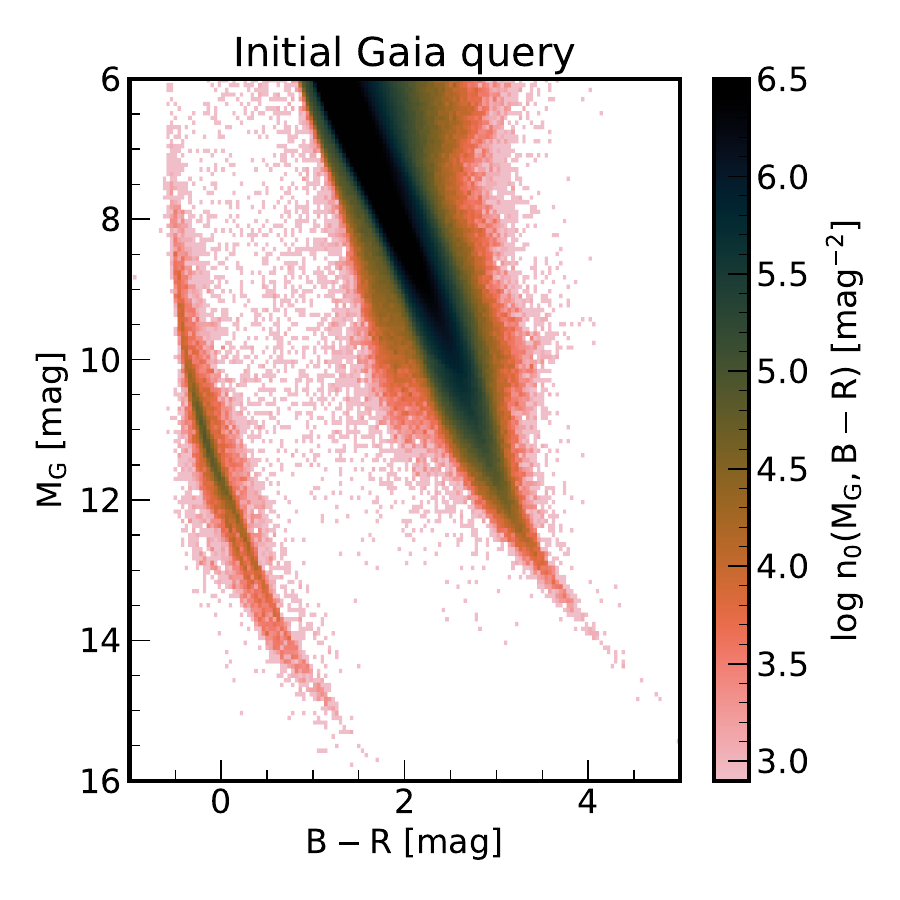}
    \includegraphics[width=0.32\linewidth]{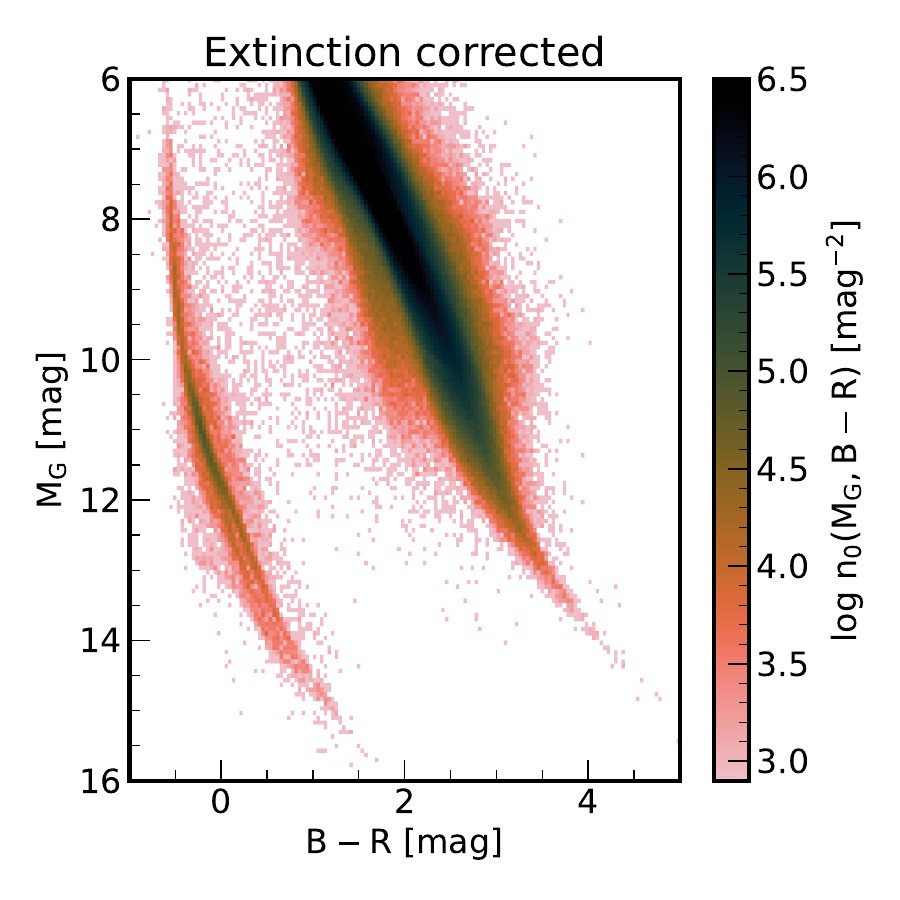}
    \includegraphics[width=0.32\linewidth]{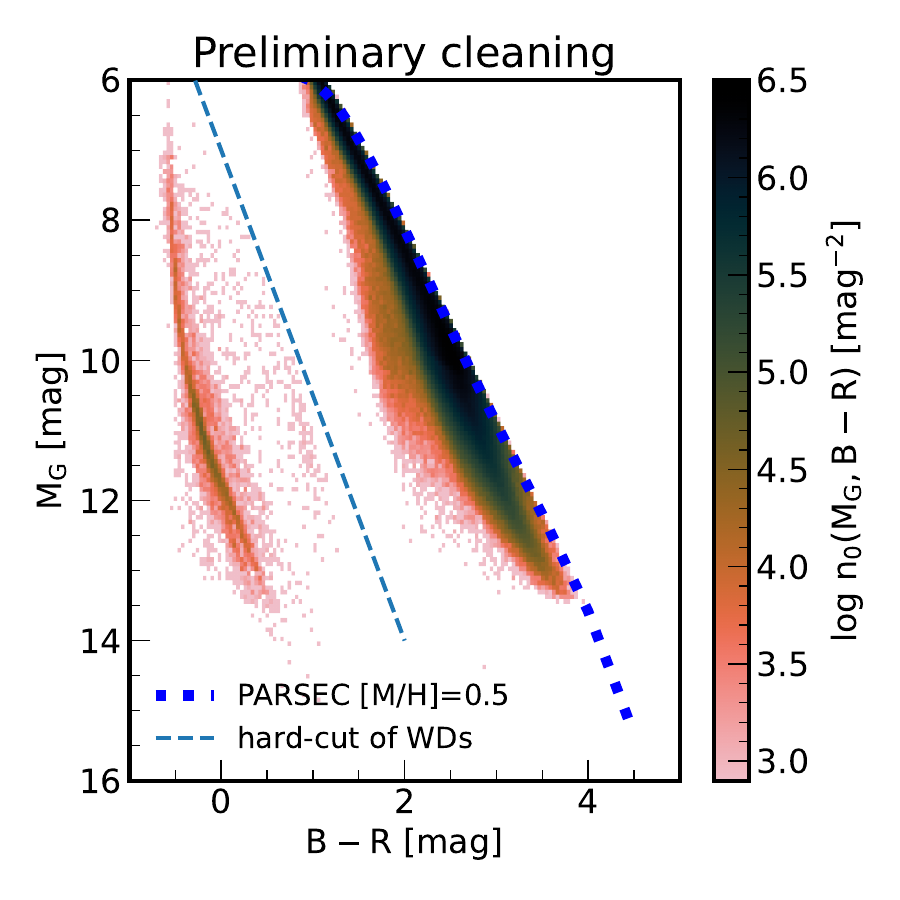}
    \caption{
    Left: Color-magnitude diagram (CMD) showing the density distribution of stars of selected \gaia\, data, with absolute G-band magnitude ($M_G$) plotted against $B_p$-$R_p$ color (B-R). 
    Middle: Extinction-corrected CMD for the stars displayed in the left panel, accounting for interstellar reddening. 
    Right: The CMD showing the distribution of stars in our sample within $\sim$300 pc that passed our initial cleaning for training the spectral emulator. 
    The blue dotted line represents a PARSEC isochrone for [M/H] $= 0.5$,  which serves as an upper boundary for selecting single main sequence stars. 
    The blue dashed line is a hard cut to select white dwarf stars. }
    \label{fig:initial query}
\end{figure*}

\subsection{From XP coefficients to sampled spectra}\label{subsec:sample_xp}

As mentioned in subsection \ref{subsec:xp}, we transform the XP coefficients back into wavelength space following the approach of \cite{zhang2023}. 
We employ the {\tt\string GaiaXPy}\footnote{http://doi.org/10.5281/zenodo.6674521} package to convert the covariance matrices of the 110 basis function coefficients into the sampled wavelength space, covering 392--992~nm with a 10~nm step. 
To ensure reliable spectral modeling, we account for covariances between fluxes at different wavelengths and apply appropriate zero-point corrections. 
The mathematical framework for handling covariance matrices and calculating goodness-of-fit statistics is presented in Appendix \ref{app:sample_xp}.

\subsection{Parallax error propagation}

We normalize all XP spectra to the flux at 10 pc using the parallax from \gaia\ DR3. 
Since the parallax measurement has uncertainties that cannot be neglected, we perform error propagation as follows:
\begin{equation}
\vec{f}_{10} = \vec{f}_{\rm obs}  \left(\frac{1}{10\varpi}\right)^2
\end{equation}
where $\vec{f}_{10}$ is the flux normalized to 10 pc, $\vec{f}_{\rm obs}$ is the observed flux at the star's distance, and $\varpi$ is the parallax in arcseconds. 
The error propagation process accounts for the uncertainty in parallax measurements when normalizing XP flux spectra to a standard distance of 10 pc.
The function applies the inverse square law to scale the observed flux ($F_d$) to its equivalent at 10 pc ($\vec{f}_{10}$). It then propagates the errors using two components:

It scales the original covariance matrix ($C_{\vec{f}{obs}}$) by the square of the scaling factor ($g_\pi$), 
\begin{equation}
C_{\vec{f}_{10}} = g_\varpi^2 C_{\vec{f}{obs}} + J J^T \sigma_\varpi^2,
\end{equation}
where $g_\varpi$ is defined as the scaling factor:
\begin{equation}
g_\varpi = \left(\frac{1}{10\varpi}\right)^2;
\end{equation}
The second variance term, $J J^T \sigma_\varpi^2$ arising from the parallax uncertainty. 
This term is computed using the Jacobian ($J$) of the scaling function with respect to parallax, multiplied by the square of the parallax error ($\sigma_\varpi$), where the Jacobian $J$ is defined as:
\begin{equation}
J = \frac{\partial g_\varpi}{\partial \varpi} \vec{f}_{\rm obs} = -\frac{2}{100\varpi^3} \vec{f}_{\rm obs}.
\end{equation}

The resulting covariance matrix ($C_{\vec{f}_{10}}$) incorporates both the scaled original uncertainties and the additional uncertainty of the parallax measurement.

\subsection{Interstellar extinction correction}\label{subsec:extinction}

Accurate extinction corrections are essential for two aspects of our work: (1) obtaining de-reddened XP spectra, and (2) determining extinction-corrected photometric labels ($M_{G0}$, (B-R)$_0$ colors) for building our forward model. 
We used the 3D dust map from \cite{Edenhofer2024}. 
Using the \texttt{dustmaps} package \citep{2018JOSS....3..695M}, we query the 3D dust map to obtain the extinction parameter $E$\footnote{The dustmap is in units of E per parsec, and can be converted to an extinction at any wavelength by multiplied the extinction curve.}
for each star. 
We then compute the fractional extinction $f_\text{ext}$ as follows:
\begin{equation}
f_\text{ext}(\lambda) = \exp(-A(\lambda) \cdot E),
\end{equation}
where $A(\lambda)$ is the interpolated extinction curve. The de-reddened XP flux $\vec{f_0}$ is obtained by:

\begin{equation}
\vec{f_0} = f_\text{ext}(\lambda) \cdot \vec{f_\text{obs}},
\end{equation}
where $\vec{f_\text{obs}}$ is the observed XP flux. We neglect the error caused by the dust map in this correction.

To compute the extinction in \gaia\ G, BP, and RP bands, we utilize their respective transmission curves $T_X(\lambda)$ and apply:
\begin{equation}
A_X = -2.5 \log_{10}\left(1 - \int (1 - f_\text{ext}(\lambda)) \cdot T_X(\lambda) \cdot d\lambda \right),
\end{equation}
where $X$ represents G, BP, or RP. 
The resulting values $A_G$, $A_{BP}$, and $A_{RP}$ are then used to correct the observed magnitudes and fluxes for interstellar extinction, allowing for a more accurate representation of the intrinsic stellar properties. 

\subsection{Training data}\label{subsec:traindata}

To mitigate the impact of interstellar dust extinction on the observed spectra, we restricted our training data to nearby stars within $\sim$333 parsecs (parallax $\varpi > 3$ milliarcseconds). 
This proximity ensures that the effects of dust attenuation are minimized, allowing us to construct a more accurate forward spectral model.
Before modeling WDMS binary systems, it is crucial to establish a clean single-star template. 
To achieve this, we performed an initial cleaning procedure on our training data, aimed at minimizing contamination from WDMS binary stars.

\subsubsection{White dwarf sample}

We used the \gaia\ DR3 white dwarf sample compiled by \cite{Garcia-Zamora2023}, which draws on the Montreal White Dwarf Database \citep{Dufour2017} as training data.
\cite{Garcia-Zamora2023} trained a Random Forest algorithm to perform spectral classification of the white dwarf population within 100 pc based on the Hermite coefficients of \gaia\ spectra.
 
We adopt \cite{Garcia-Zamora2023}'s WD sample as our initial white dwarf sample, which includes various spectral subtypes: e.g., hydrogen-rich (DA) and hydrogen-deficient (non-DA) WD stars.
From the initial sample of 95,533 stars, we select 34,988 with an average signal-to-noise ratio (S/N) $> 5$.

Additionally, we applied a strict cut on the color-magnitude diagram (CMD) position, as shown in Fig~\ref{fig:initial query}. 
Although our signal-to-noise (S/N) threshold of >5 ensures good data quality, it excludes many white dwarfs fainter than $M_G < 14$. 
This is because such faint WDs typically have apparent G magnitudes $>18$, where XP spectra quality degrades \citep{DeAngeli2022}. 
Despite potentially missing some faint WDs, maintaining this minimum S/N requirement is crucial for reliable spectral analysis.

\subsubsection{Main-sequence star sample}

Our initial selection encompasses all M stars within approximately 333 parsecs. 
To guide our selection process, we employ PARSEC isochrones corresponding to an age of 1 Gyr and an overall metallicity of [M/H]$=0.5$. 
By retaining only the stars positioned below this isochrone in the CMD, we removed most potential MS-MS binaries and young stellar objects (YSOs) from our sample.
Second, we compare the integrated flux between two wavelength ranges to identify and exclude WDMS binary systems. 
To build pure spectral templates for both main-sequence stars and white dwarfs, we eliminate potential WDMS binary contaminants from our initial training sample. 
Specifically, we calculate the ratio of fluxes: $F_{492-582}$ (492--582 nm) to $F_{392-482}$ (392--482 nm). 
A WDMS binary typically shows stronger flux in the bluer wavelength range due to the contribution from the main sequence companion. 
Therefore, we classify an object as a WDMS candidate if:
\begin{equation}
    \frac{F_{492-582}}{F_{392-482}} < 1,
\end{equation}
this step is necessary because some WDMS binary stars exhibit blue excess in the BP band due to the white dwarf companion. 
This cleaning step is crucial for ensuring that our templates represent the true spectral characteristics of single stars.
Finally, we select stars with an average S/N $> 20$. 
After applying these criteria, we obtain 1,610,396 stars' XP spectra as our initial training sample, as shown in the right panel of Fig.~\ref{fig:initial query}. 

\section{Spectral model}\label{sec:method}

Our spectral modelling for MSWD binaries consists of two main steps: (1) building two data-driven forward model for single-star spectra, one for white dwarfs and one for main-sequence stars; 
and (2) fitting each observed spectrum with both single-star and binary models to identify observed systems that are not well described by a single-star model, but significantly better described by a binary-star model.

\subsection{Single-star spectral model}

\begin{figure*}
    \centering
    \includegraphics[width=\linewidth]{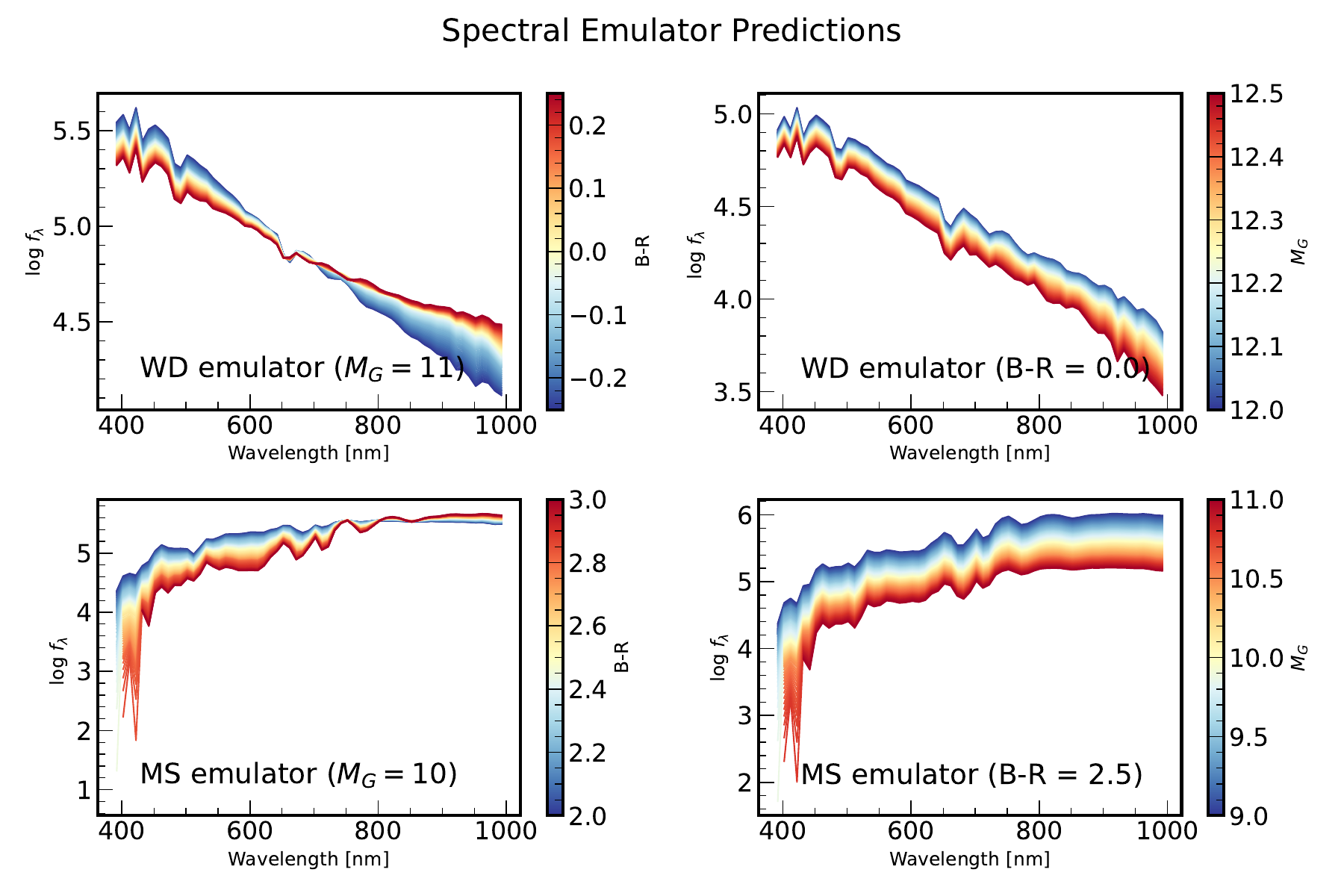}
\caption{Demonstration of our NN spectral emulators for white dwarfs (top panels) and main sequence stars (bottom panels). 
Each panel shows the predicted XP spectra with different stellar parameters, where the x-axis is the wavelength and the y-axis is the logarithmic flux ($10^{-20}~{\rm W}~{\rm m}^{-2}~{\rm nm}$) normalised to 10 pc. 
Left panels: Spectra at fixed absolute G magnitude ($M_G$) with different B-R color. 
Right panels: Spectra at fixed B-R color with varying $M_G$. }
    \label{fig:single_sed_model}
\end{figure*}

We first build a single-star emulator, i.e. a data-driven generative model for the XP spectra of single stars. 
This emulator is based on a feedforward neural network (NN) architecture that predicts the flux at a given wavelength as a function of stellar labels ($\boldsymbol{\theta}$).
Traditional stellar labels such as temperature and surface gravity may not be optimal for a NN, when considering the dichotomous nature of WD and MS stars. 
Instead, we adopt vary basic photometric quantities as our stellar labels:
$\boldsymbol{\theta} = (B - R, M_G, M_R)$
where $B - R$ is the dust-corrected color, and $M_G$ and $M_R$ are the (dust-corrected) absolute magnitudes in \gaia's G and R bands, respectively.

The neural network comprises five hidden layers, each containing 64 neurons activated by Rectified Linear Units (ReLU) function. 
This configuration allows the model to capture non-linear relationships between the stellar labels and the resulting spectra. 
We train the network using the \texttt{Adam} optimizer with a learning rate of $10^{-4}$ and a batch size of 16,384 samples, which balances computational efficiency with stable convergence.

This NN takes the stellar labels $\boldsymbol{\theta}$ as input and outputs a vector representing the XP spectrum:
$f(\lambda) = \text{NN}(\boldsymbol{\theta}; \boldsymbol{w})$,
where $f(\lambda)$ is the predicted flux at wavelength $\lambda$, and $\boldsymbol{w}$ represents the weights and biases of the NN.
We train the NN using a mean-squared-error loss function. 

The single-star model is a forward model that maps labels $\boldsymbol{\theta}$ to XP spectra. 
The free parameters of the fitting function (the NN weights and biases) are determined by optimizing the model on the training set described in Section \ref{subsec:traindata}. Implicitly, this assumes that each of the two training sets consists only of single stars (either MS or WD), or at least is dominated by single stars. 
We tackle this issue below by iterative ``cleaning'' of the sample used to build these single-star emulators.

The result of building these two single-star emulators is shown in Figure~\ref{fig:single_sed_model}. 
The emulated spectra not only have physically sensible shapes, but also show some of the strong absorption lines. The left two sub-panels show how the emulated spectra vary with color (at a given absolute magnitude) for MS and WD; the right panels show how the spectra vary with absolute magnitude at a given color.

\subsection{Binary-star spectral model}

To model the spectra of WDMS binary systems, we considered the weighted sums of spectra drawn from the two just-described single-star models, rather than training a separate binary model. 
In this approach, each of the two single-star models is characterized by a set of labels: $\boldsymbol{\theta}_{\text{WD}}$ for the white dwarf and $\boldsymbol{\theta}_{\text{MS}}$ for the main sequence stars;
$\boldsymbol{\theta}_{\text{WD}} = (B - R)_{\text{WD}}, M_{G,\text{WD}}, M_{R,\text{WD}}$
$\boldsymbol{\theta}_{\text{MS}} = (B - R)_{\text{MS}}, M_{G,\text{MS}}, M_{R,\text{MS}}$

The binary spectrum is then modeled as a straightforward sum of the two-component spectra, each component created by its neural network model for single stars:

\begin{equation}
    \mathbf{f}_{\text{binary}}(\lambda) = \text{NN}(\boldsymbol{\theta}_{\text{WD}}; \mathbf{w}) + \text{NN}(\boldsymbol{\theta}_{\text{MS}}; \mathbf{w})
\end{equation}

This approach allows us to efficiently model binary systems without the need for additional training and all within the framework of our original neural network architecture.  We use this binary-star model to identify systems that are better described by a binary (rather than a single) configuration.

\subsection{Spectral model fitting and validation}

We used least-squares optimization with a Trust Region Reflective (TRF) algorithm \citep{coleman1994centering}, when fitting both the single-star emulators and the binary models to the observed spectra. 
For single-star fits, the stellar parameters $\boldsymbol{\theta} = (B-R, M_G, M_R)$ are allowed to vary across the combined parameter space of both WD and MS stars. This way we get a direct comparison between WD and MS models to determine which stellar type better represents the data.
For binary fits, we restrict the stellar parameters to ranges that are physically motivated for each of the components: for white dwarf components, we require $-0.5 \leq (B-R)_{\rm WD} \leq 2$ and $6 \leq M_{G,\rm WD} \leq 15$, while for MS components, we set $1 \leq (B-R)_{\rm MS} \leq 4$ and $6 \leq M_{G,\rm MS} \leq 15$. 
This ensures that the optimization explores only realistic WDMS binary configurations, avoiding non-physical combinations while looking for the best-fit solution.

After performing single-star and binary fittings, we derive the \singlechi and \binarychi values for each star. 
Since our goal is to build emulators from pure single-star spectra, 
To further refine our emulator, we perform additional cleaning on the training data and re-train the emulators. 
This is achieved by performing three iterations:
In each iteration, we retain only stars that meet two criteria:
(1) \singlechi - \binarychi < 0, indicating that the single star fit is better than the binary fit, and (2) \singlechi / DOF < 20, ensuring a good fit quality, where DOF is the degree of freedom.
We aim to eliminate contamination of WDMS binaries from our sample by repeating this process three times.

\begin{figure*}
    \centering
    \includegraphics[width=\linewidth]{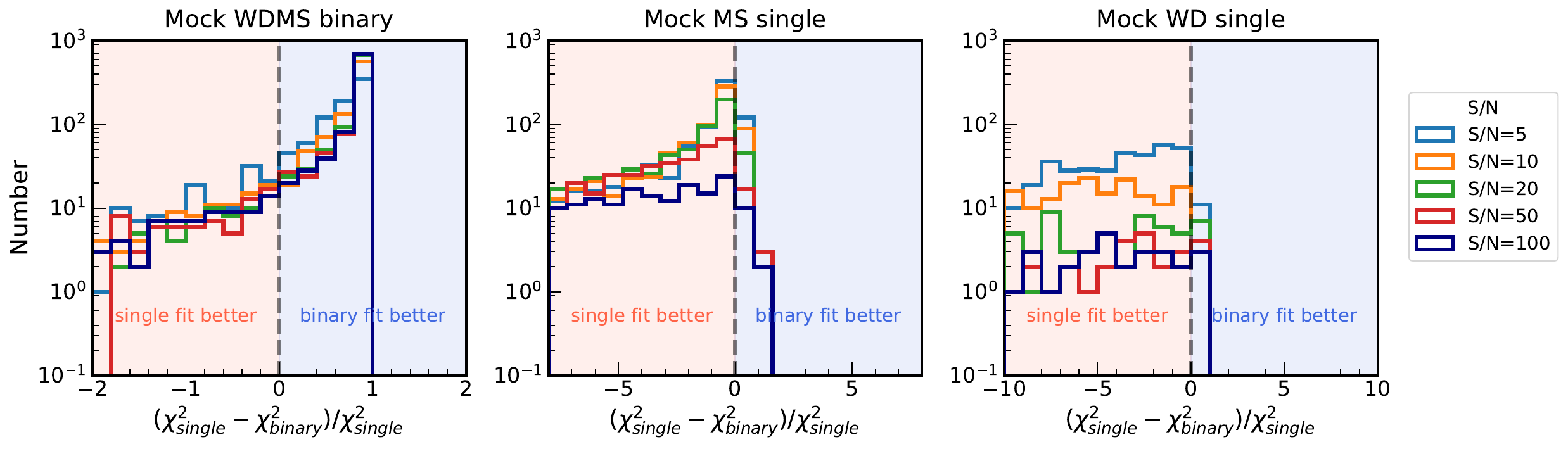}
    \caption{
    Comparison of the re-normalized $\chi^2$ differences between single and binary star models for three different mock datasets: (a) Mock WDMS binary, (b) Mock MS single, and (c) Mock WD single. The x-axis represents the normalized difference in chi-squared values, (($\chi^2_{\text{single}} - \chi^2_{\text{WDMS}})$ / $\chi^2_{\text{single}}$), and the y-axis shows the number of occurrences on a logarithmic scale. The histograms are color-coded by signal-to-noise ratio (S/N) values of 5, 10, 20, 50, and 100. 
    The percentage accuracy of correctly identifying the better fit (single or binary) is indicated in the legend for each S/N value. 
    Vertical dashed lines at (x=0) separate regions where single or binary models provide a better fit. 
}
    \label{fig:chi2_compare}
\end{figure*}

\begin{figure}
    \centering
    \includegraphics[width=0.8\linewidth]{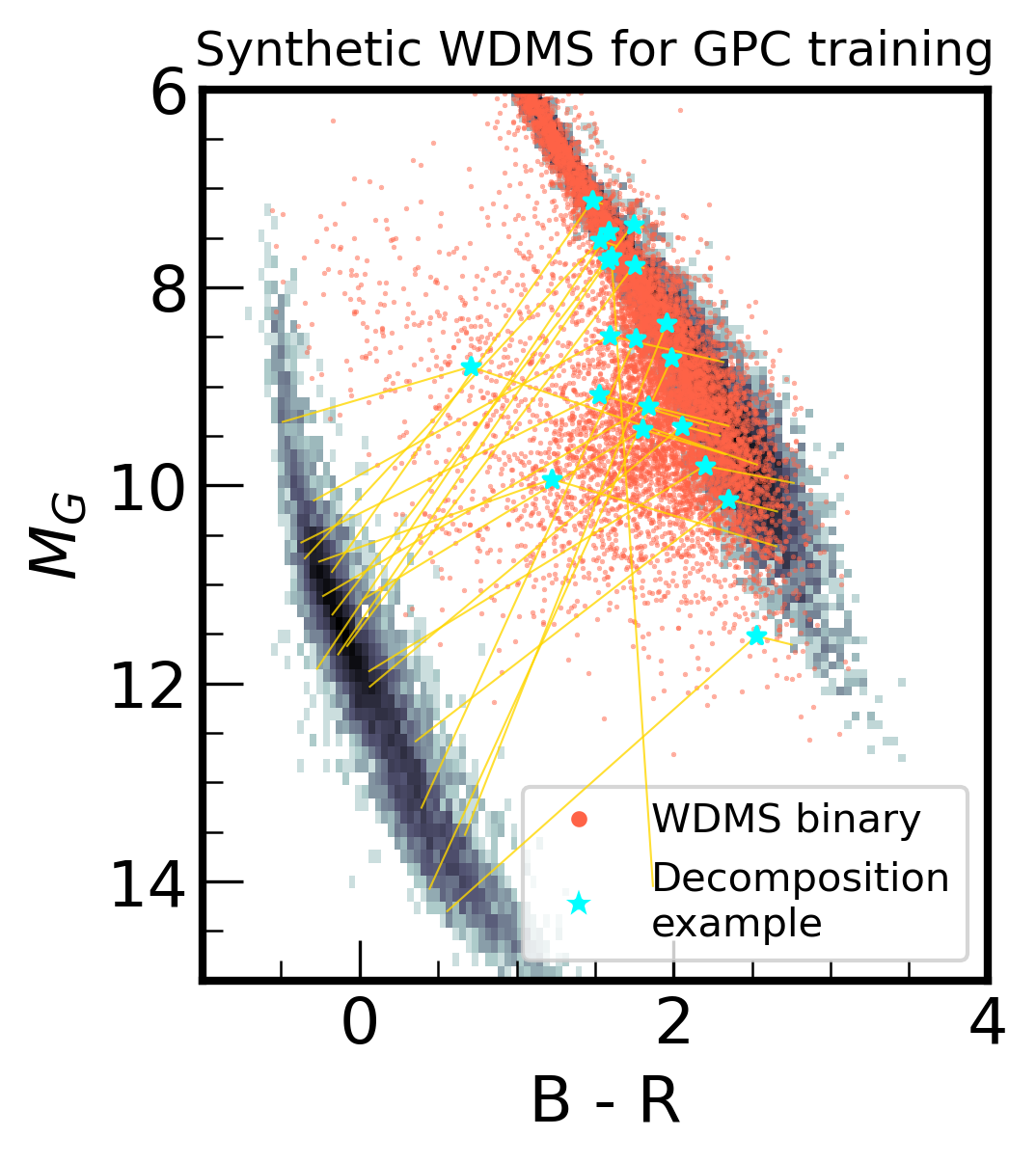}
    \caption{CMD illustrating the mock  data of single main-sequence (MS) stars, single white dwarfs (WD), the synthetic WDMS binary systems. 
    The gray 2-d histograms background represent the white dwarf sequence,
   and the main sequence stars. The red stars indicate synthetic WDMS binaries. 
    The yellow lines connect the WD and MS components of individual binary systems for synthetic WDMS. }
    \label{fig:HRD_binary_trainGPC}
\end{figure}

To validate our method, we first train a Normalizing Flow \citep{pmlr-v37-rezende15}
on the label space ($M_G$, $B-R$, $M_R$) of our cleaned single-star samples. 
We then draw samples from this trained flow to generate realistic stellar parameters, which we input into our trained emulators to create 10,000 mock spectra each for single MS and WD stars (see Appendix~\ref{app:nf} for details). This approach ensures that our mock spectra follow the intrinsic distribution of stellar parameters observed in our training sample.

We then create 10,000 mock WDMS binary spectra by combining the two types of single-star model fluxes, and add Gaussian noise corresponding to different signal-to-noise ratios (S/N). 
This yields a total validation set of 30,000 spectra at each S/N level.

We fit each spectrum twice, first with a single-star model and subsequently with a binary model, which gives us $\chi^2_{\mathrm{single}}$ and $\chi^2_{\mathrm{WDMS}}$ values. The classification performance is evaluated using the renormalized $\chi^2$ difference:

\begin{equation}
    \Delta \chi^2_{\mathrm{norm}} = \frac{\chi^2_{\mathrm{single}} - \chi^2_{\mathrm{WDMS}}}{\chi^2_{\mathrm{single}}},
\end{equation}
where negative values indicate a better single-star fit and positive values favor the binary model. 
At S/N = 20, the method achieves high accuracy in identifying single stars (89\% for MS, 99\% for WDs) but lower accuracy (67\%) for binary systems, implying the need for more sophisticated detection methods beyond just the spectral fitting approach.

\begin{table*}
\centering
\caption{Selection Criteria for $\chi^2$-selected WDMS Binaries}
\begin{tabular}{p{0.35\textwidth} p{0.55\textwidth}}
\hline\hline
Category & Selection Criterion \\
\hline
Better binary fit & $\chi^2_{\mathrm{single}} - \chi^2_{\mathrm{WDMS}} > 0$ \\
Quality Cut & $\frac{\chi^2_{\mathrm{single}}}{\mathrm{DOF}} < 20$ and $\frac{\chi^2_{\mathrm{WDMS}}}{\mathrm{DOF}} < 20$ \\
Comparable Luminosity & $|M_{R,\mathrm{MS}} - M_{R,\mathrm{WD}}| < 2.5$ \\
                      & $|M_{G,\mathrm{MS}} - M_{G,\mathrm{WD}}| < 2.5$ \\
WD Component Location & To the left of the dotted line in Fig.~\ref{fig:hrd_wdmsfit} \\
\hline
\multicolumn{2}{p{0.9\textwidth}}{\textit{Note:} The magnitude difference limit ensures components differ by less than a factor of 10 in brightness.} \\
\hline
\end{tabular}
\label{tab:selection_criteria}
\end{table*}

The limitation has inspired us to consider approach to identifying WDMS binaries which employs two complementary methods. 
First, we used the $\chi^2$ fitting to identify a high-purity sample of WDMS binaries where both components contribute visibly to the XP spectra. 
We found this method to be reliable, but we also found that it often misses systems where one component dominates the flux. 
Therefore,  we developed the Gaussian Process Classification (GPC) technique, described in Section~\ref{sec:gpc}, that incorporates multiple spectral features beyond $\chi^2$ statistics. 
This hybrid approach allows us to identify a more complete sample of WDMS binaries, particularly systems containing faint WDs and bright MS stars that we found challenging to detect through $\chi^2$ fitting alone. 

\begin{figure*}[hbt!]
    \centering
    \includegraphics[width=\linewidth]{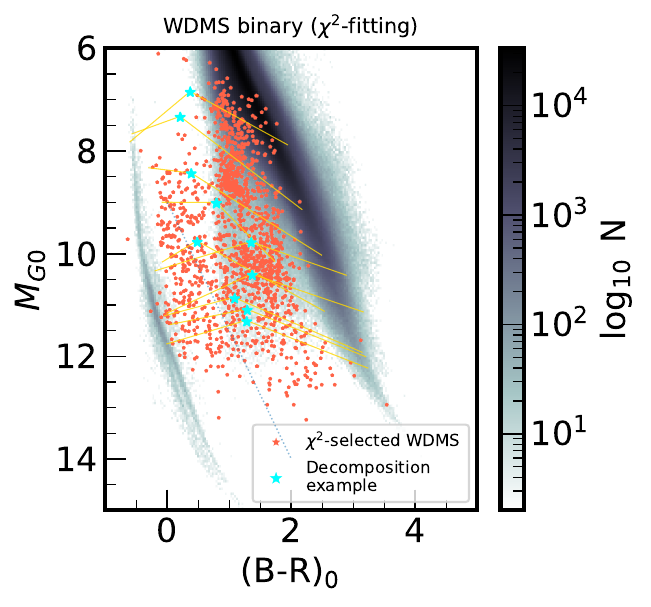}
    \caption{CMD showing the distribution of WDMS binary candidates identified through $\chi^2$ fitting of XP spectra (red dots). 
    The background grayscale shows the density distribution of all stars, with the scale bar indicating the logarithmic number of stars per bin. 
    The x-axis shows the de-reddened \gaia\ color (B-R)$_0$, while the y-axis shows the absolute G magnitude ($M_{G0}$) after extinction correction. 
    The 10 cyan stars are examples of the $\chi^2$ selected WDMS binaries, the yellow lines show the best fit spectral decomposition of a WD and an MS by XP spectra.
    The binary candidates primarily occupy the region between the main sequence (right) and white dwarf sequence (left), as expected for composite systems.
    The blue dotted line denotes the requirements of the WD component of a binary model fit, which should lie to the left of this line as in the WD regime.}
    \label{fig:hrd_wdmsfit}
\end{figure*}

\subsection{Identification of WDMS binaries through binary-model fitting}\label{subsec:chi2_wdms}

\begin{figure*}[hbt]
    \centering
    \includegraphics[width=0.85\linewidth]{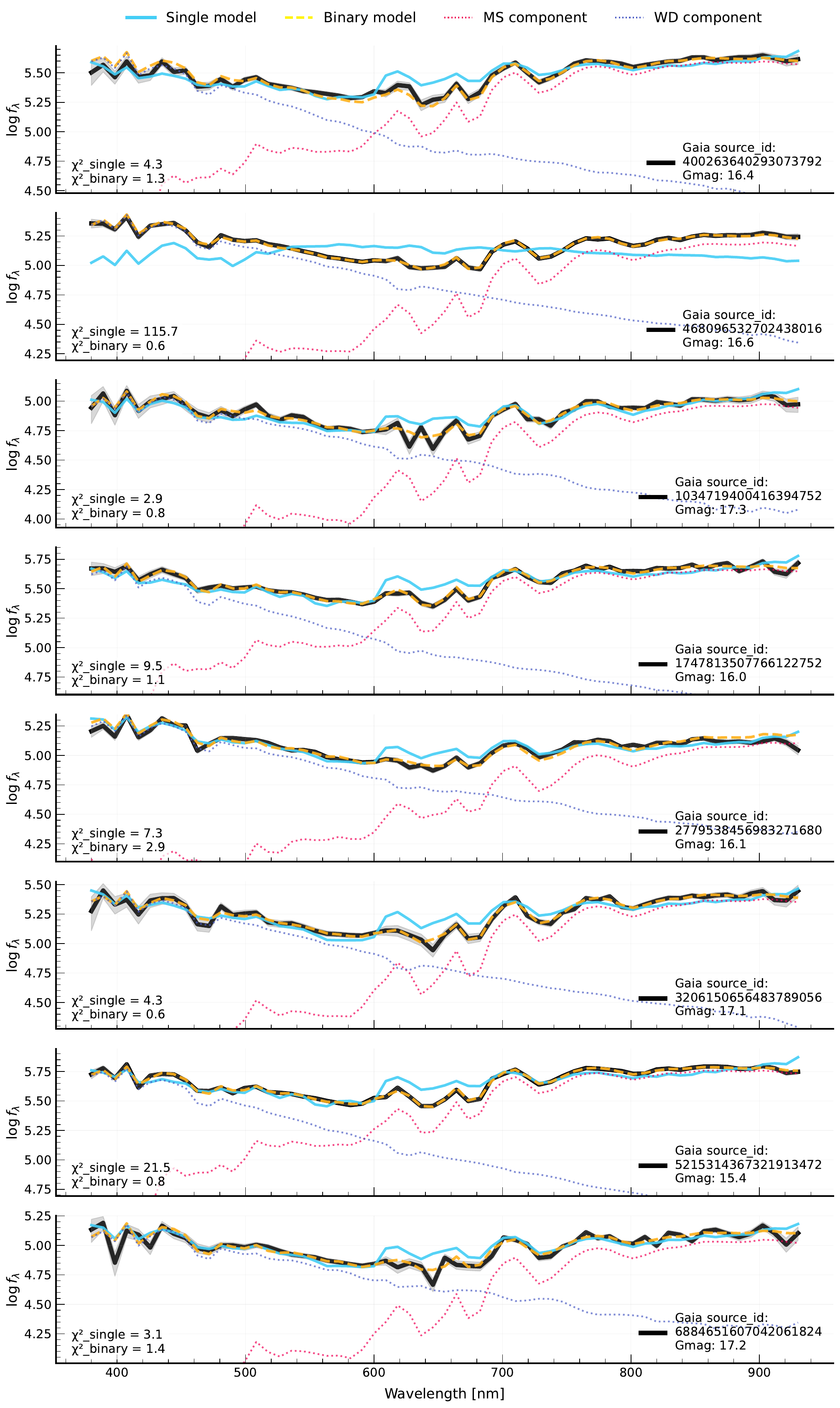}
    \caption{
    \gaia\ XP spectra and best-fitting models for eight of ten WDMS binaries selected using the $\chi^2$ fitting method are shown as examples in Fig.~\ref{fig:hrd_wdmsfit}. 
    In each panel, the black line represents the observed flux, with the grey shading indicating flux error. 
    The cyan solid line represents the best-fit single model, while the orange dashed line shows the best-fit binary model. 
    The pink dotted line corresponds to the MS component of the binary model, and the blue dotted line represents the WD component of the binary fit.}
\label{fig:fittingresult}
\end{figure*}

To identify reliable WDMS binary candidates from spectral fitting, we have to specify how good or bad the single-star fit is  (in a $\chi^2$-sense), and how much better the binary model fit is. To do this, we implemented several selection criteria based on both the quality of the fit and the physical properties of the fitted components, spelled out in Table~\ref{tab:selection_criteria}:

First, we require that both the single- and binary-star model fits have reasonable quality, with total $\chi^2$ values (computed across all 61 spectral pixels) less than 20, and we require the binary fit to be better than the single fit.
This ensures that the spectral models provide meaningful representations of the observed data. 
 However, if the single-star fit has $\chi^2 > 20$, the candidate is excluded, as this reflects a poor model match to the observed spectrum, even with a good binary fit. 
 The binary fit must not only meet the $\chi^2 < 20$ threshold, but also show significant improvement over the single-star fit to ensure the selection of robust binary candidates. 

Second, we impose constraints on the relative luminosities of the WD and MS components. 
As shown in Table \ref{tab:selection_criteria}, we require that the magnitude differences between the components in both \gaia\ G and R bands cut to be less than 2.5 magnitudes. 
This criterion ensures that the luminosity ratio between components is within an order of manitude, as we have found that binary fits become unreliable when one component is dramatically fainter than the other, given the typical signal-to-noise ratios of the \gaia\ XP spectra.

Third, we verify that the fitted WD component has physically plausible parameters by requiring it to lie in the expected region of the CMD for white dwarfs. 
We implement this using a simple linear cut in the CMD defined by points
$(B-R = 0, M_G = 9)$ and $(B-R = 2, M_G = 14)$ as shown in Fig.~\ref{fig:hrd_wdmsfit}.
Objects are only retained if their fitted WD component falls below this line in the CMD.
A system must satisfy all three of these criteria to be classified as a candidate binary WDMS. 
This conservative approach boosts the purity of the sample selection, at the expense of excluding some genuine WDMS systems, in particular those where one component dominates the observed spectra.

In Fig.~\ref{fig:hrd_wdmsfit}, we show the CMD positions of all 1,649 WDMS binary systems that satisfy our selection criteria (Table~\ref{tab:selection_criteria}). 
The candidates predominantly occupy the region between the main sequence and the white dwarf cooling sequence, as expected for the photometric properties of unresolved WDMS binaries. 
This distribution supports the effectiveness of our selection methodology. Of course, this fitting not only identifies the systems as WDMS binaries, but also provides an estimate of the individual colors and luminosities for both components.

\section{A Gaussian Process Classifier for WDMS binaries}\label{sec:gpc}

Although our $\chi^2$-based selection criteria yield a highly pure sample of WDMS binaries, this approach has limitations. 
The choice of $\chi^2$ thresholds is inherently arbitrary and hence sub-optimal, as it should vary with the relative brightness of the binary components (or S/N) and CMD position
(see Fig.~\ref{fig:chi2_compare}). 
This is problematic for systems where one component outshines the other, leading to an incomplete sample. 
Strict cuts $\chi^2$ will not recognize genuine WDMS systems even if they fall just outside our chosen thresholds.
To overcome these limitations, we develop a novel approach in the following subsection, employing a Gaussian Process Classifier. 
This method provides a more flexible and statistically robust framework for binary classification, allowing us to incorporate multiple features beyond simple $\chi^2$ statistics and automatically handle varying detection thresholds across different regions of parameter space.

Therefore, we introduce and apply Gaussian Process Classification (GPC) \citep{Rasmussen2006Gaussian}, 
a probabilistic extension of Gaussian processes designed for classification tasks while providing robust uncertainty quantification. 
Unlike methods such as logistic regression that impose linear decision boundaries, GPC can learn complex, nonlinear class separations without requiring manual feature engineering. As shown in Appendix \ref{app:GPC}, GPC operates by modeling a latent function through a Gaussian process, which is then transformed via tapering function to generate class probabilities.

The advantage of GPC lies in its mathematically principled approach to uncertainty quantification in classification problems. 
This is achieved through multiple mechanisms: a kernel function that captures similarity between data points, automatic modulation of prediction confidence based on training data proximity, and probability estimates that naturally reflect uncertainty in sparsely sampled regions of the parameter space. 
Although the non-Gaussian likelihood inherent to classification precludes exact solutions, GPC employs the Laplace approximation to maintain computational tractability while preserving accurate uncertainty estimation. 
For a mathematical treatment of these concepts, we refer the reader to the appendix \ref{app:GPC}.

\subsection{Training the Gaussian Process Classifier}

To create our training dataset, we first generate synthetic spectra using our single-star and binary-star spectral emulators developed in Section \ref{sec:method}, specifically
10,000 single WD spectra and 10,000 single MS spectra on the XP wavelength grid, as shown in Figure \ref{fig:HRD_binary_trainGPC}.

We chose the 10,000 synthetic WDMS binary, 10,000 single WD spectra, and 10,000 single MS spectra to ensure model training across all stellar types. 
We randomly split each population 80\%-20\%, to get a training set of 24,000 spectra (8,000 of each type) and a test set of 6,000 spectra (2,000 of each type). This split maintains class balance in both sets while ensuring our validation metrics, shown in Figure \ref{fig:HRD_binary_trainGPC}'s CMD, reflect the model's true generalization performance across the full range of stellar populations.
This training set is then used to optimize the parameters of the GPC model, while the test set is held out to provide an unbiased assessment of classification performance. 
This GPC model training involves tuning hyperparameters to minimize classification error, with details of the optimization process provided in Appendix~\ref{app:GPC}. 

As a second aspect of the training, we fit both single-star and binary-star models to all mock spectra, as described above, to obtain $\chi^2_{\rm single}$ and $\chi^2_{\rm WDMS}$ values for each. 
We then use this $\chi^2$ difference as an 
initial classification feature, supplemented by features from XP spectra.
This choice balances classification performance and computational efficiency. 
Reducing them to synthetic colors lowers dimensionality while retaining essential information for classification.
We supplement this with synthetic photometry computed from the XP spectra in SkyMapper \citep{Keller2007}'s passbands ($v$, $g$, $r$, $i$, and $z$). 
After testing various combinations of colors, we identify four key discriminating features: the renormalized $\chi^2$ difference ($\Delta \chi^2_{\mathrm{norm}}$) and three SkyMapper colors ($v-r$, $v-g$, and $i-z$).
The training set consisted of mock single stars and mock binary systems, with their positions plotted in multi-dimensional color-magnitude space as shown in Fig.~\ref{fig:corner_mocktrain} in Appendix~\ref{app:GPC}.

For binary classification, we implement a GPC using a Radial Basis Function (RBF) kernel with a kernel size of 0.1.
Prior to training, we apply standard normalization to all features by transforming each feature $x \rightarrow \frac{x - \mu}{\sigma},$
where $\mu$ is the mean and $\sigma$ is the standard deviation of each feature.
The choice of the hyperparameter is elaborated in Appendix~\ref{app:GPC}. 
This configuration enables the GPC to capture both the global structure and local variations in our feature space.

\subsection{Validation of the Gaussian Process Classifier}

\begin{figure*}
    \centering
    \includegraphics[width=\linewidth]{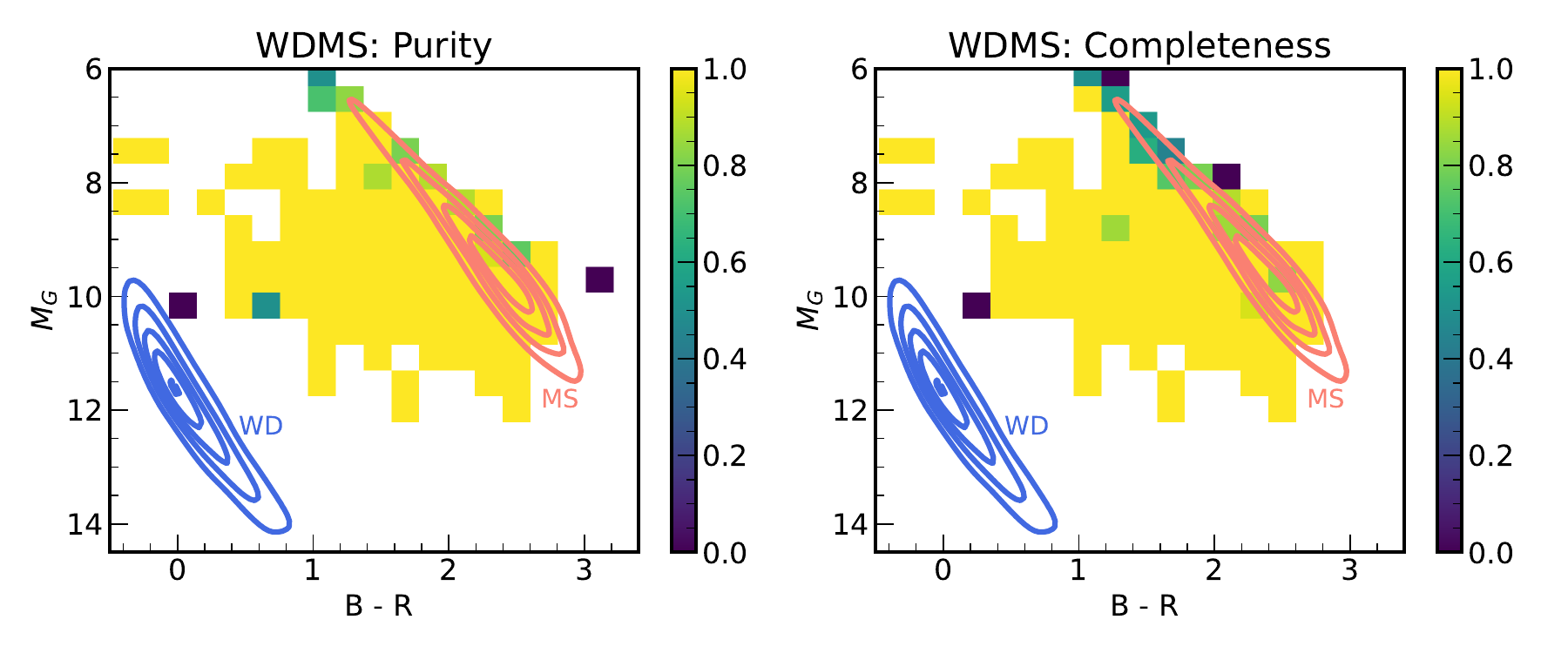}
        \caption{
CMDs illustrating the classification performance metrics of the GPC method for identifying WDMS binaries within the validation dataset, which is based on synthetic spectra.
        \textit{Left:} Purity (precision) map showing the fraction of true WDMS binaries among all objects classified as WDMS in each CMD pixel. \textit{Right:} Completeness (recall) map showing the fraction of correctly identified WDMS binaries among all true WDMS binaries in each CMD pixel. 
        The color scale ranges from 0 (purple) to 1 (yellow), where yellower regions indicate better performance.
        The blue and red contours represent the density distribution of individual white dwarf (WD) and main sequence (MS) stars, respectively, in CMD space. 
        Both metrics demonstrate robust performance (yellow regions) in the characteristic WDMS locus, with some expected degradation (darker colors) in regions where WDMS binaries overlap with the single-star main sequence populations.}
    \label{fig:gpc_cmd}
\end{figure*}
Our performance metrics for GPC classification are derived from a validation dataset containing synthetic WDMS systems, created by randomly pairing WD and MS stars observed within 300 pc. 
While the color-magnitude positions of the individual binary components are drawn from real observations, this random pairing assumption may not fully capture the true binary formation physics, where certain types of WDs could preferentially form binaries with specific MS populations due to evolutionary effects. 
Therefore, the actual classification performance for observed WDMS systems might differ from our estimates, particularly in CMD regions, where binary formation physics introduces correlations between WD and MS properties.

To evaluate our classification performance, we focus on two key metrics.
The first metric is \emph{purity}, also known as precision, which measures the reliability of our positive classifications by calculating the proportion of true WDMS binaries among all objects classified as WDMS. 
The second metric is \emph{completeness}, also known as recall, which  quantifies our detection efficiency by measuring the proportion of correctly identified WDMS binaries among all true WDMS systems in our sample.


Both performance measures are expected to vary as a function of CMD position, and they do, as illustrated in Figure~\ref{fig:gpc_cmd}.
These color-coded CMD maps reveal that both metrics achieve optimal performance in the characteristic WDMS locus between the single WD and MS populations (shown as blue and red contours respectively). 
The classification precision and purity shows expected degradation in regions where WDMS binaries overlap with single-star populations, particularly at the boundaries of MS distributions. 

\subsection{Application of Gaussian Process Classification to WDMS  Classification}\label{subsec:gpc_wdms}

\begin{figure*}
    \centering
    \includegraphics[width=\linewidth]{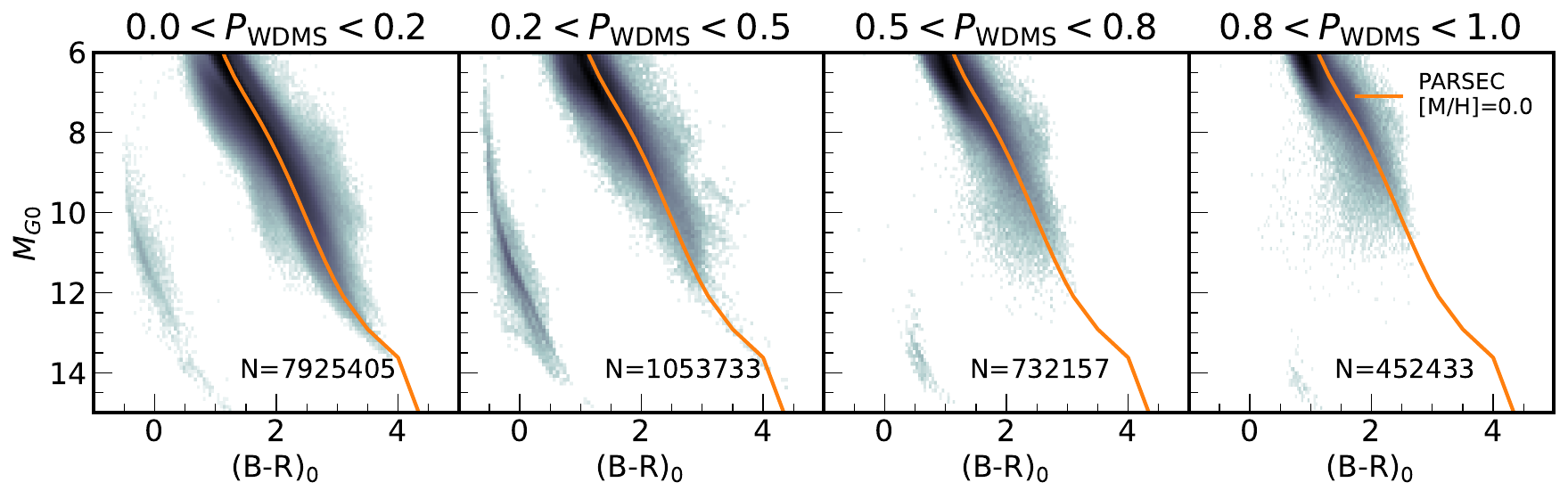}
    \caption{CMD showing the classification of stars into single and binary systems by the Gaussian Process Classifier (GPC). 
    The four panels show different probability ranges for binary classification:  $0.0 < P_{\rm WDMS} < 0.2$, $0.2 < P_{\rm WDMS} < 0.5$, $0.5 < P_{\rm WDMS} < 0.8$, $0.8< P_{\rm WDMS}<1.0$.
    The orange line represents the PARSEC isochrone with [M/H]=0.0. 
    The distribution reveals that stars falling between the main sequence and white dwarf regions are more likely to be classified as binaries, consistent with expectations for WDMS binary systems.}
    \label{fig:camd_prob_binary}
\end{figure*}
\begin{figure*}[hbt!]
    \centering
    \includegraphics[width=\linewidth]{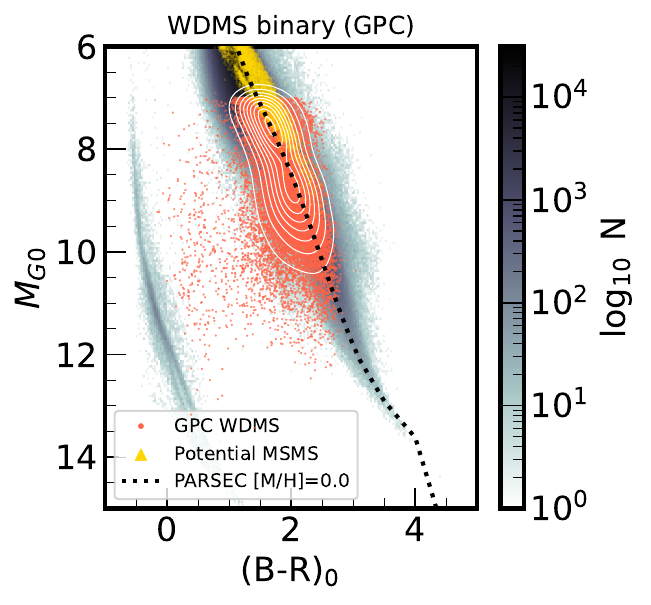}
\caption{CMD showing WDMS binary candidates identified by our GPC method (red dots). 
Yellow triangles denote possible MSMS binaries.
The white contours show the density distribution of the GPC-WDMS candidates.
The background grayscale shows the density distribution of all stars in our sample, with the colorbar indicating the logarithmic number of stars per bin, which reflects a refined sample from Figure \ref{fig:camd_prob_binary}.
}
    \label{fig:HRD_gpc}
\end{figure*}

We now employ this classification approach, in which the GPC draws on the $\Delta\chi^2$ output from the spectral fitting, to our complete dataset described in Section~\ref{sec:method}. 
In a preparatory step, we correct for the interstellar extinction present in the \gaia\ photometric measurements (G, B, R bands) and the XP spectra as shown in \ref{subsec:extinction}.

Utilizing these extinction-corrected data, we compute the SkyMapper color indices ($v-r$, $v-g$, $i-z$) from the XP spectra to serve as classification features.
Subsequently, we perform fits on each extinction-corrected XP spectrum using both single-star and binary-star models to calculate the re-normalized $\chi^2$ difference between them. 
This computed $\chi^2$ difference and all SkyMapper color indices then constitute the feature set utilized for the GPC classification algorithm to assign a binary probability to each star. 
Figure~\ref{fig:camd_prob_binary} illustrates the CMD for each GPC probability bin. 
Consistent with expectations, stars positioned between the MS and WD sequences exhibit the highest probabilities of being binary systems.

If we adopt only probability a threshold of $P_{\rm WDMS}>0.8$, we identify 452,433 WDMS binary candidates. However, we then apply additional quality criteria to ensure the reliability of our final sample:
\begin{enumerate}
    \item At least one of our two spectral models must yield a good fit to the XP spectra (either \binarychi\ /DOF or \singlechi\ /DOF less than 20), ensuring the validity of our spectral analysis;
    \item The extinction in G-band ($A_G$) must be less than 0.5 mag to minimize contamination from highly reddened sources;
    \item The absolute magnitude after extinction correction ($M_{G0}$) must be fainter than 7. Stars brighter than this threshold would show significantly larger luminosity differences if they contain a WD component, as most white dwarfs have absolute magnitudes fainter than 8. This cut was validated through our cross-matching with {\it GALEX} data \citep{Martin2005} (also see  the discussion in Section~\ref{diss:galex}).
    \item Exclude stars as MSMS binaries if $\chi^2_{\rm MSMS}$ is smaller than both $\chi^2_{\rm single}$ and $\chi^2_{\rm WDMS}$. These excluded stars are represented as yellow dots in Fig.~\ref{fig:HRD_gpc} (see the discussion in Subsection~\ref{diss:galex}). Stars are classified as MSMS binaries (flagged as \texttt{flag$_{\rm MSMS}$}) if they satisfy the following two conditions:
    (1) the difference $\chi^2_{\rm single}$/DOF - $\chi^2_{\rm MSMS}$/DOF is greater than 10, and (2) the value $\chi^2_{\rm MSMS} - \chi^2_{\rm WDMS}$ is less than 0.
    The terms $\chi^2_{\rm single}$, $\chi^2_{\rm MSMS}$, and $\chi^2_{\rm WDMS}$ are defined as follows: The $\chi^2_{\rm single}$ represents the reduced $\chi^2$ value for the fit assuming a single-star model, which can correspond to a MS or a WD. The $\chi^2_{\rm MSMS}$ represents the $\chi^2$ value for the fit assuming an MSMS (Main Sequence + Main Sequence) binary system, describing the fit quality of the MSMS binary model. Finally, the $\chi^2_{\rm WDMS}$ corresponds to the reduced $\chi^2$ value for the fit assuming a WDMS binary system.
    \item The normalized $\chi^2$ difference, $(\chi^2_{\rm single} - \chi^2_{\rm WDMS})/\chi^2_{\rm single}$ must exceed -1 to ensure that the binary fit is either better than the single fit, or at least not much worse.
\end{enumerate}
These quality cuts eliminate the vast majority of initial candidates, but still leave a vast final WDMS candidate sample, consisting of 27, 606 WDMS systems, as shown in Figure~\ref{fig:HRD_gpc}. Compared to selection methods based solely on $\chi^2$ statistics (Figure~\ref{fig:hrd_wdmsfit}), the GPC classification approach identifies a much larger population of candidates near or even on the main sequence. 

These additional candidates likely represent genuine WDMS systems comprising a bright main-sequence star and a faint white dwarf companion that had proven challenging to identify with $\chi^2$ fitting methods.
We publish the GPC-WDMS catalog (\ref{tab:cat}) online: \url{https://zenodo.org/records/14411003} \citep{li_2024_14411003}.

\begin{table*}\label{tab:cat}
\centering
\caption{Description of columns in the WDMS binary catalog  in \cite{li_2024_14411003}}
\begin{tabular}{p{4cm}p{2cm}p{8cm}}
\hline\hline
Column Name & Data Type & Description \\
\hline
\texttt{source\_id} & \texttt{int64} & Unique \gaia\ DR3 source identifier \\
\texttt{ra} & \texttt{float64} & Right ascension by \gaia\ DR3 (deg) \\
\texttt{dec} & \texttt{float64} & Declination by \gaia\ DR3 (deg) \\
\texttt{parallax} & \texttt{float64} & Parallax (mas) \\
\texttt{parallax\_over\_error} & \texttt{float64} & Parallax divided by its standard error \\
\texttt{ruwe} & \texttt{float64} & Renormalized unit weight error\\
\texttt{phot\_g\_mean\_mag} & \texttt{float64} & Mean magnitude in \gaia\ G band \\
\texttt{phot\_bp\_mean\_mag} & \texttt{float64} & Mean magnitude in \gaia\ BP band \\
\texttt{phot\_rp\_mean\_mag} & \texttt{float64} & Mean magnitude in \gaia\ RP band \\
\texttt{bp\_rp} & \texttt{float64} & BP-RP color \\
\texttt{chi2\_diff} & \texttt{float64} & Difference in $\chi^2$ between single MS or WD and WDMS binary model fits \\
\texttt{chi2\_diff\_renorm} & \texttt{float64} & Renormalized $\chi^2$ difference \\
\texttt{chi2\_single} & \texttt{float64} & $\chi^2$ of single MS or WD model fit \\
\texttt{chi2\_binary} & \texttt{float64} & $\chi^2$ of WDMS binary model fit \\
\texttt{bp\_rp\_single} & \texttt{float64} & BP-RP color from single MS or WD fit \\
\texttt{abs\_g\_single} & \texttt{float64} & Absolute G magnitude from single MS or WD fit \\
\texttt{abs\_r\_single} & \texttt{float64} & Absolute RP magnitude from single MS or WD fit \\
\texttt{bp\_rp\_wd} & \texttt{float64} & BP-RP color of WD component \\
\texttt{abs\_g\_wd} & \texttt{float64} & Absolute G magnitude of WD component \\
\texttt{abs\_r\_wd} & \texttt{float64} & Absolute RP magnitude of WD component \\
\texttt{bp\_rp\_md} & \texttt{float64} & BP-RP color of MS component \\
\texttt{abs\_g\_md} & \texttt{float64} & Absolute G magnitude of MS component \\
\texttt{abs\_r\_md} & \texttt{float64} & Absolute RP magnitude of MS component \\
\texttt{prob\_binary} & \texttt{float64} & Probability that the source is a WDMS binary system \\
\texttt{flag\_binary\_fit} & \texttt{bool} & Quality flag for WDMS binary model fit (0=good, 1=poor) \\
\texttt{flag\_binary\_mag\_diff} & \texttt{bool} & Flag indicating large magnitude difference between components\\
\texttt{flag\_single\_fit} & \texttt{bool} & Quality flag for single MS or WD fit (0=good, 1=poor) \\
\texttt{flag\_wdmsfit\_in} & \texttt{bool} & Mask indicating whether the WD companion locate in WD sequence \\ 
  &  &  (0=not in, 1=in) \\
\texttt{chi2\_mass\_mh} & \texttt{float} & $\chi^2$ value for the single MS fit \\
\texttt{chi2\_mass\_mh\_binary} & \texttt{float} & $\chi^2$ value for the MSMS binary fit \\
\texttt{q\_binary} & \texttt{float} & Mass ratio ($q$) of the secondary to the primary in the MSMS binary fit \\
\texttt{chi2\_diff\_mh\_mass} & \texttt{float} & Difference in $\chi^2$ between single MS and MSMS binary model fits \\
\texttt{flag\_MSMS} & \texttt{bool} & Flag indicating MSMS binary classification (1=MSMS binary) \\
\hline
\end{tabular}
\label{tab:columns}
\end{table*}

\subsection{Incompleteness and Training Data Limitations}\label{subsec:incompleteness}
The Gaussian Process Classifier (GPC) is a discriminative model, and hence the classification of WDMS systems is influenced by the distribution of the training data \citep{Hogg2024}. 
Our GPC relies on the distribution of its training data, which was constructed by randomly pairing observed white dwarfs and main sequence stars within 300 pc. 
However, this assumption of random pairings may not reflect the true physical distribution of WDMS binary components. 
More broadly, the discriminative nature of our GPC implies that its classification boundaries are fundamentally shaped by the priors encoded in the training data. 
Any systematic differences between these assumed priors and the true underlying population will impact both the completeness and reliability of our classifications. While our validation against GALEX UV data (see Section~\ref{diss:galex}) confirms the high reliability for the systems we do identify, the catalog's completeness is likely to vary across parameter space in ways tied to these training set limitations.
This illustrates an inherent challenge in using discriminative models like GPC for population studies -- the method can separate populations with properties similar to the training data, but may miss systems that evolved through pathways not well represented in the training.

A second key limitation arises from the quality constraints of our training data. Due to the signal-to-noise limitations of XP spectra, our white dwarf sample is restricted to objects brighter than $M_{G0}$ = 14 (see Fig.~\ref{fig:initial query}). 
This introduces a bias against cool, old white dwarfs with cooling ages greater than 3 Gyr.

\section{Discussion}\label{sec:discussion}

We have shown that the low-resolution \gaia\ XP spectra are a powerful tool for identifying WDMS binaries. Yet, we found that even this approach still faces serious challenges, as binaries' combined spectral energy distributions will vary dramatically depending on the properties of their components. In this section, we validate our catalog and analyze its properties through three complementary investigations.

First, we verify our classification using the modest subset of systems where UV photometry is available, since the inevitably hotter white dwarf will almost always affect the UV colors.

Second, we will validate our method through the comparison with existing catalogs, cross-matching our sample with previously identified WDMS binaries from SDSS and LAMOST spectroscopic surveys (in subsection\ref{subsec:catalog_comparison}). 
We also examine agreement with binaries identified through independent techniques, including SED-based and astrometric methods that probe different regions of parameter space.

Third, we present a statistical analysis of our $\chi^2$-selected WDMS binary sample (in Subsection~\ref{subsec:physical_properties}). In this context, we examine the distribution of the isochrone-inferred masses for the WD component, which reveals an excess of lower-mass white dwarfs, compared to single white dwarfs. If this is not a selection effect, it may provide insight into binary evolution pathways.

Finally, we investigate selection effects and sources of incompleteness in our sample (subsection \ref{subsec:incompleteness}),  as they are crucial for understanding the limitations of our catalog and its implications for binary population studies.

\subsection{Binary validation using UV Data}\label{diss:galex}

\begin{table*}
\caption{Numbers of stars that pass different selection criteria.}
\begin{tabular}{p{10cm}p{3cm}p{2cm}}
\hline
Selection criteria & Number of stars  \\
\hline
All star sample & 10,163,728 \\
\texttt{Chi2}: High-purity sample ($\chi^2$ selection) & 1,649  \\
\texttt{QC}: GPC sample with quality cuts (qc)$^a$ & 55,050  \\
\texttt{FaintQC}$^b$: GPC sample with qc and $M_{G0} > 7$,   \texttt{flag$_{\rm MSMS}$}=0 & 27,444  \\
GPC sample with qc $\times$ GALEX, $M_{G0} > 7$ & 3,404 \\
GPC sample with qc $\times$ GALEX, $M_{G0} > 7$ & 2,416  \\
+show GALEX NUV excess & & \\
\hline
\end{tabular}
\begin{flushleft}
\small
$^a$Quality cuts include: flag\_binary\_fit=False, flag\_single\_fit=False, chi2\_diff\_renorm$>$-1, $A_G<0.5$\\
$^b$Recommendation of complementarity between purity and completeness
\end{flushleft}
\label{tab:sample_stats}
\end{table*}

\begin{figure*}
    \centering
    \includegraphics[width=0.99\linewidth]{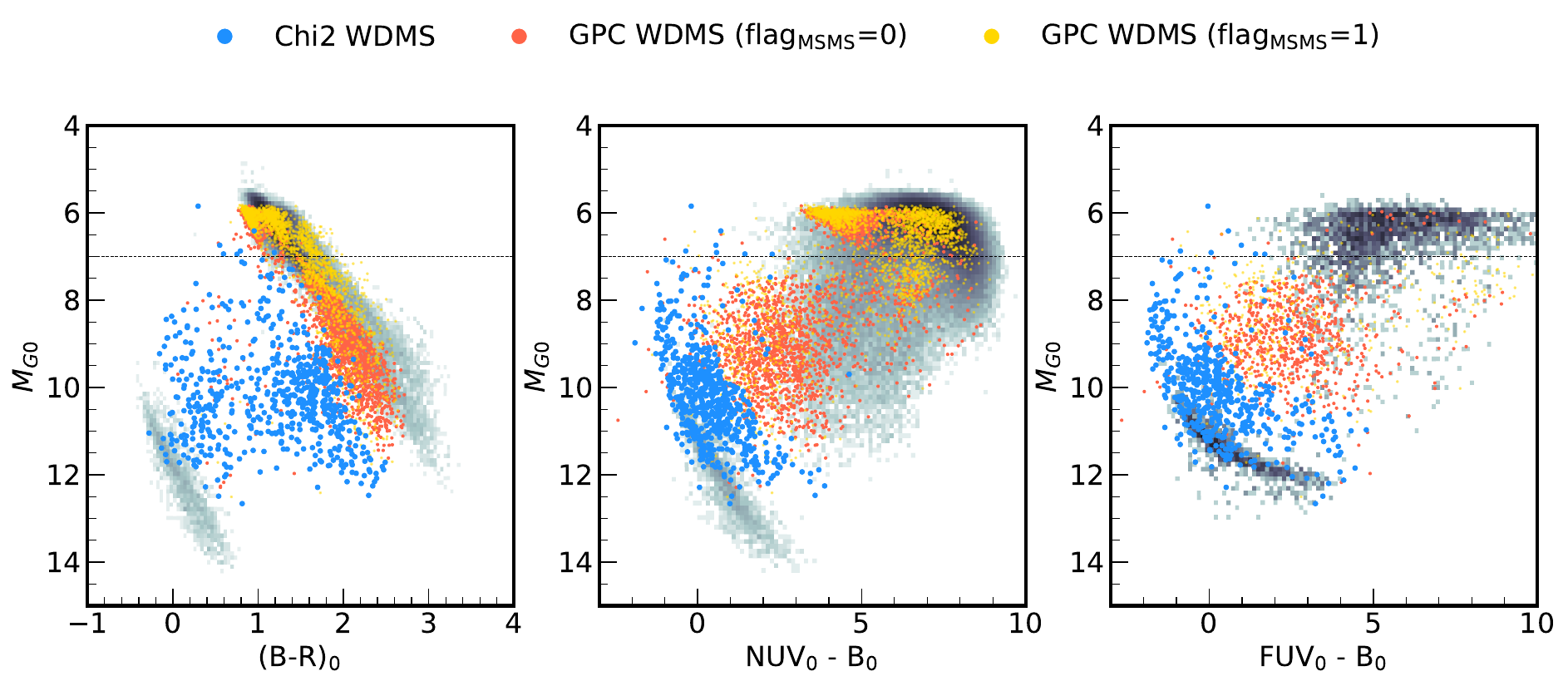}
   \caption{
   Comparison of WDMS binary candidates selected by different methods in various CMDs. 
   Left panel: \gaia\ CMD showing absolute G magnitude versus de-reddened B-R color. 
   Middle and right panels: GALEX-Gaia CMD using NUV and FUV bands, respectively. 
   In all panels, blue dots represent WDMS binaries identified through $\chi^2$-fitting of their XP spectra (\texttt{Chi2} sample in Table~\ref{tab:sample_stats}), while red dots show WDMS candidates identified by our Gaussian Process Classifier (GPC) (\texttt{QC} sample) but missed by the \texttt{Chi2} sample. 
   Yellow dots mark possible MSMS binaries (\texttt{flag$_{\rm MSMS}$}=1), mostly brighter than the dotted line at $M_{G0} = 7$. 
   The underlying grey points show the single stars, as defined by the GPC with $P_{\rm WDMS}<0.2$.
   }
    \label{fig:galex}
\end{figure*}

Here we use the near- and far-UV photometry from the GALEX mission \citep{Martin2005} to verify WDMS binary candidates identified through our two different methods: the $\Delta\chi^2$ statistics and GPC. 
Recall that we applied quality cuts (including flag criteria, extinction limits, and $\chi^2$ thresholds) to ensure the GPC classifications are based on reliable input data. 
The UV colors of these candidate WDMS binaries are shown in Figure~\ref{fig:galex}. 
Grey points show the color-color positions of (single) WDs and MS stars. The colored symbols show the WDMW binaries, with $\chi^2$-identifies systems in blue, and GPC-identified systems in red.

In the \gaia\ CMD, $\chi^2$-identified candidates predominantly occupy the expected gap between the main sequence and white dwarf loci, as shown in Figure~\ref{fig:hrd_wdmsfit}. Some of the GPC-identified candidates lie in the same CMD gap. However, many of GPC-identified candidates lie along the main sequence itself, as illustrated in Fig.~\ref{fig:gpc_cmd}. For these, the optical CMD position by itself is inconclusive.

To validate all these candidates, we now consider the subsample -- among the $\chi^2$ and GPC identified binary samples, and the single stars sample ($P_{\rm WDMS}<0.2$) -- that can be cross-matched with \textit{GALEX} UV data \citep{Martin2005} .
Among the GPC-selected candidates that passed our quality cuts, 6,484 sources have GALEX NUV photometry, 1,182 stars have both NUV and FUV measurements.
We use an emiprical relation \citep{ZhangR2023} to estimate the extinction in the \textit{GALEX} near-UV (NUV) band.
The extinction is given by:
\begin{align}
A_{\rm NUV} &\approx \frac{R_{\rm NUV}}{R_{\rm G}}A_{\rm G} \approx 3.085 A_{\rm G}, \\
A_{\rm FUV} &\approx \frac{R_{\rm FUV}}{R_{\rm G}}A_{\rm G} \approx 2.95 A_{\rm G},
\end{align}
where $A_{\rm NUV}$, $A_{\rm FUV}$, $A_{\rm G}$ represent the extinction in the NUV, FUV and $G$ bands, respectively, and $R_{\rm NUV}$, $R_{\rm FUV}$, and $R_{\rm G}$ are their corresponding extinction coefficients.

As shown in the middle panel of Fig.~\ref{fig:galex}, most WDMS candidates identified via $\chi^2$ stand out in the near-UV CMD on account of their (blue) UV colors; this is of course unsurprising, given that they already stand out in the \gaia\ CMD.  The near-UV CMDs are more interesting for GPC-identified WDMS binaries missed by spectral-fitting selection. 
Approximately 50\% of these occupy the region in the near-UV CMD  between the WD cooling sequence and main sequence in the NUV$_0$-B$_0$ versus $M_{G0}$ diagram. 
These systems show distinct NUV excess compared to main sequence stars, but are not as blue as typical white dwarfs. 
In contrast, the $\chi^2$-selected WDMS binaries have NUV magnitudes similar to single white dwarfs, clustering along the WD sequence. 
This intermediate NUV brightness of many GPC-identified systems provides evidence that they are genuine WDMS binaries missed by the spectral fitting methods.

However, the remaining $\sim$50\% of GPC-identified candidates lie in the same region as main sequence stars in the NUV$_0$-B$_0$ versus $M_{G0}$ diagram. 
Without FUV data (as shown in Fig.~\ref{fig:galex}), it is difficult to confirm the binary nature of these systems. 
Additionally, most of them lie in a clump of candidates at $M_{G0} \sim 6$, which are possibly main-sequence main-sequence (MSMS) binaries or a triple system involving MSMS binaries and a WD third component.
The nature of these ($2615$) systems (\texttt{flag$_{\rm MSMS}$}=True, defined in Subsection~\ref{subsec:gpc_wdms}) 
is unclear: they may be possibly MSMS binaries not included in the training set and may arise from applying the classifier to real data, where it otherwise performs exceptionally well. But we need to consider most of these misclassification as WDMS binaries.

This analysis summarized in Table~\ref{tab:sample_stats}, shows the progression from our initial all-sky sample to increasingly refined subsets of WDMS binary candidates, with UV data providing crucial independent validation of our classifications.

\subsection{Comparison with existing literature}\label{subsec:catalog_comparison}

We now validate our WDMS binary candidates through comparisons with existing samples, which identified using different methodologies: spectroscopic identification from SDSS and LAMOST surveys, SED fitting of \gaia\ EDR3 data within 100 pc, and astrometric detection using \\gaia\ DR3 orbital solutions. 
These catalogs complement our work by probing different regions of parameter space: spectroscopic surveys excel at identifying systems with comparable component luminosities, SED fitting is sensitive to the overall energy distribution. 
This comparison allows us to assess the reliability of our detection method.

\begin{figure*}[hbt!]
    \centering
    \includegraphics[width=0.49\linewidth]{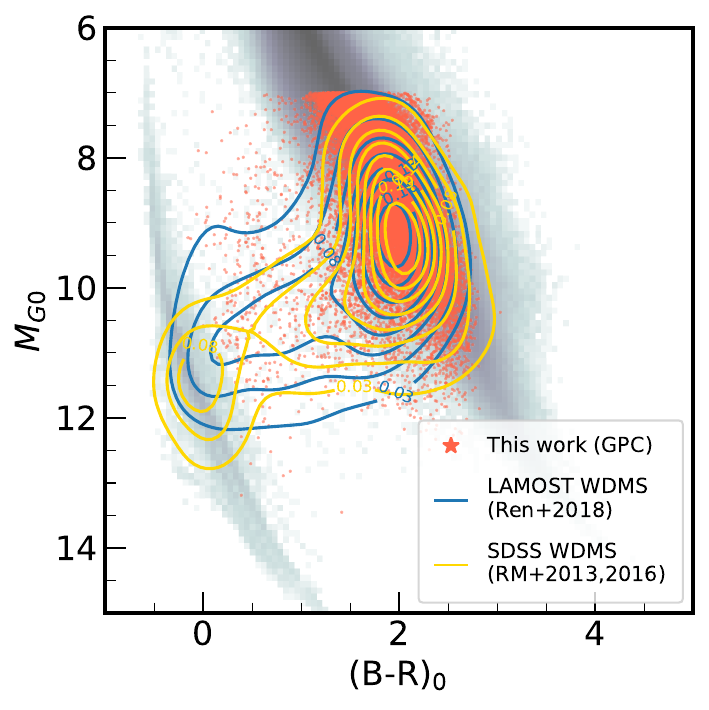}
    \includegraphics[width=0.49\linewidth]{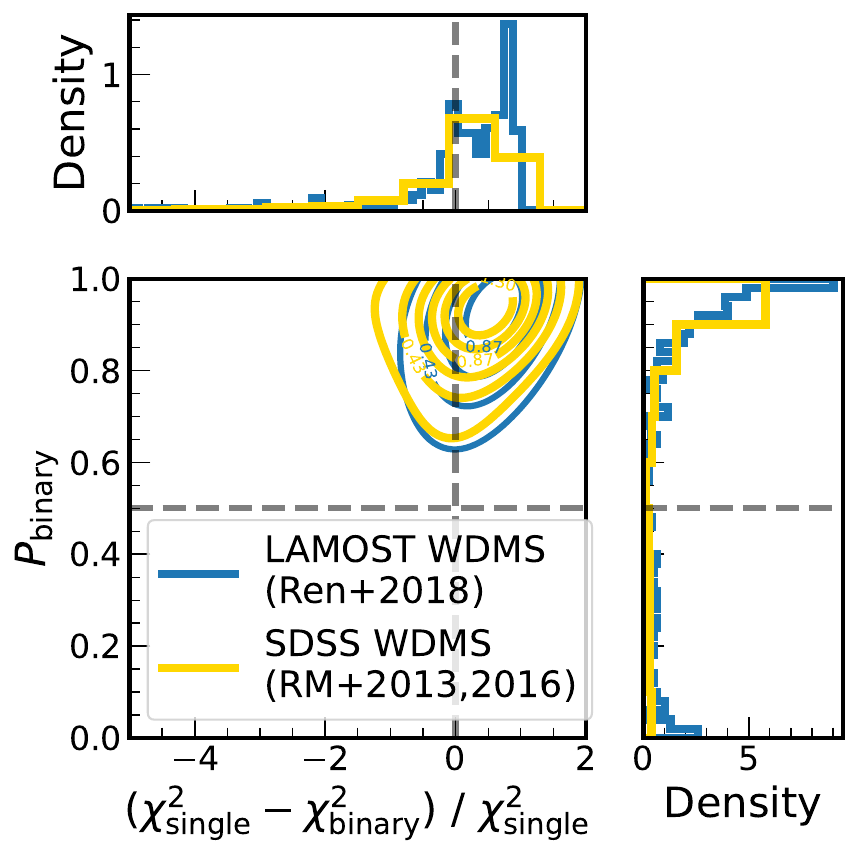}
    \caption{Left: CMD comparing WDMS binary samples from different surveys. 
    Red dots show our GPC-identified WDMS from \gaia\ XP,  the blue contour represents LAMOST DR5 WDMS binaries \citep{Ren2018}, and yellow contour shows SDSS DR12 WDMS systems \citep{Rebassa2013,Rebassa2016}.
    The background shows the density of all stars in our sample. 
    Right: The 2-D KDE distribution of the LAMOST and SDSS WDMS live in the coordinates of renormalized $\chi^2 $ difference and $P_{\rm WDMS}.$
    }
    \label{fig:HRD_gpc_lm_sdss}
\end{figure*}

\subsubsection{Spectroscopic binary}

Behind all the results derived analysing both wide WDMS binaries and PCEBs relies a tremendous observational effort dedicated to identifying large and homogeneous (i.e. selected under the same criteria) samples of such systems. 
So far, some of the largest catalogs of spectroscopic WDMS binaries have been obtained from the SDSS and LAMOST Survey (\citealt{Ren2014, Ren2018}). 
We compare with two major previous spectroscopy catalogs: the LAMOST DR5 sample from \citet{Ren2018} containing 1,121 WDMS binaries with resolution of R $\sim$ 1800, and the SDSS catalog from \citet{Rebassa2013,Rebassa2016}, which identify $\sim 3000$ WDMS through optical spectra with resolution of R $\sim$ 2000. 

From the SDSS catalog \citep{Rebassa2013,Rebassa2016}, 1,757 out of their 4,182 WDMS systems meet our basic \gaia\ astrometric and photometric selection criteria (subsection~\ref{subsec:data_selection}). 
Of these, 936 stars have XP spectra available, but only 678 stars have B mag brighter than 18, with 494 showing good model fits (defined as \singlechi$<20$ or \binarychi$<20$),
For the LAMOST sample of 1,121 WDMS binaries, 324 systems pass both our \gaia\ selection criteria and have reliable XP spectral fits.

Figure~\ref{fig:HRD_gpc_lm_sdss} shows the color-magnitude diagram (CMD) positions of our GPC-selected sample (red points) compared to the LAMOST (1,121 systems) and SDSS (1,757 systems) WDMS binaries, independent of XP spectral fit quality.
The LAMOST and SDSS samples show a concentration around $(B-R)_0 \sim 2$ and $M_{G0} \sim 9$. 
The SDSS sample extends to fainter magnitudes and includes more WD-dominated systems that occur near the white dwarf cooling sequence, a region underrepresented in our XP sample.

Figure~\ref{fig:HRD_gpc_lm_sdss} compares the two-dimensional distribution of the renormalized $\chi^2$ difference, defined as $(\chi^2_{\rm single} - \chi^2_{\rm WDMS})/\chi^2_{\rm single}$, against the binary probability ($P_{\rm WDMS}$) determined by GPC. 
The majority of the LAMOST and SDSS WDMS samples fall in the region where $P_{\rm WDMS}>0.8$ and $\chi^2$ difference$>0$, validating the effectiveness of our classification method. This high recovery rate of known WDMS systems demonstrates the robustness of our GPC-based approach. 
The systems that disagree with the GPC classification primarily lie along the white dwarf sequence, corresponding to the WDMS systems that are underrepresented in our work.

\subsubsection{SED-Identified WDMS Binaries in the Literature}

We further cross-matched our catalog with the work of \cite{Rebassa2021}, who identified 112 unresolved WDMS binaries within 100 pc using spectral energy distribution (SED) fitting of \gaia\ EDR3 data. 
Of these 112 WDMS, 70 have B magnitudes brighter than 18 and have XP spectra. 
Among the 70 systems that match the magnitude limits of our catalog, approximately 84\% have $P_{\rm WDMS} > 0.8$, indicating agreement between the two approaches. 
Of these 70 systems, 44 are common sources shared between our GPC catalog and the sample presented in \cite{Rebassa2021}.
Among the 26 missing sources, 21 lie in the WD sequence that our method does not identify, as shown in Fig.~\ref{fig:HRD_gpc_lm_sdss}. 
Two missing sources have $M_G$ fainter than 13, an absolute magnitude range absent from both our GPC catalog and the synthetic WDMS training data.

\subsubsection{Validation using astrometric binaries}
\citet{Shahaf2024} applied an astrometric triage technique based on the astrometric mass ratio function (AMRF) derived from \gaia\ DR3 astrometric orbits to identify systems with compact (or dark / low luminosity) companions. 


We select stars by applying several criteria to ensure a well-characterized sample. Primary mass $M_1 < 1.2 M_\odot$ and red excess probability $< 0.56$ initially yield 3,114 stars. 
Of these, 2,804 stars satisfy our quality criteria: \texttt{parallax\_over\_error}$>10$, available continuous XP spectra, B-band magnitude $<18$, and absolute magnitude $M_G < 7$. 
Among these stars, only 18 are included in our GPC sample. 

The limited overlap between these two samples highlights differences in selection biases and the underlying physical principles of the respective detection methods.
\citet{Shahaf2019} assume the luminosity ratio between the binary components likely containing a compact object, implying that the white dwarf (WD) contributes no light to the system. 
The AMRF method is effective in identifying such systems based solely on gravitational effects, making it a complementary technique to our spectrophotometric approach, which can only detect WDMS binaries where the WD does contribute to the observed light. 
The complementary nature of these methods is crucial. Combining both astrometric and spectrophotometric approaches provides a more complete census of the WDMS binary population.

\subsection{An excess of low-mass white dwarfs in WDMS}\label{subsec:physical_properties}

\begin{figure}
    \centering
    \includegraphics[width=0.49\linewidth]{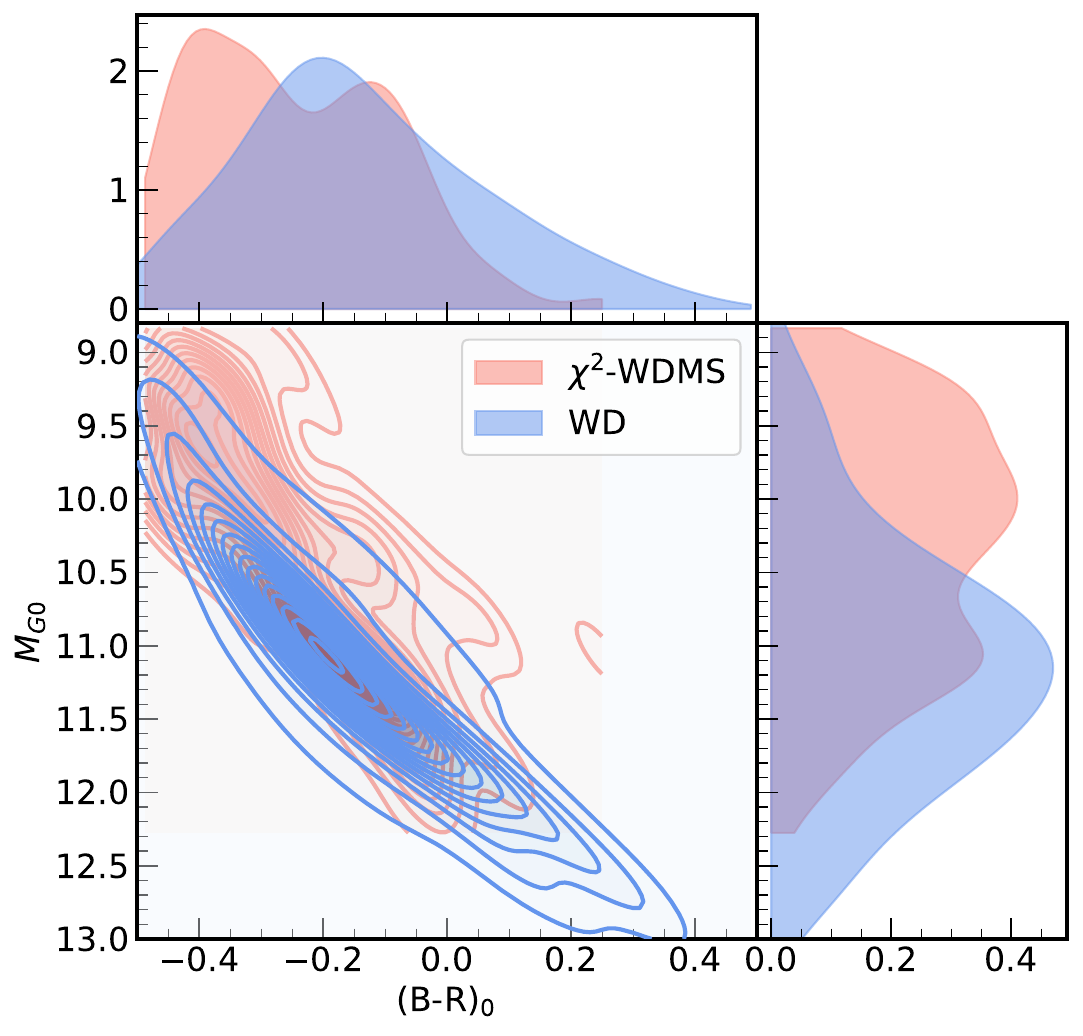}
     \includegraphics[width=0.49\linewidth]{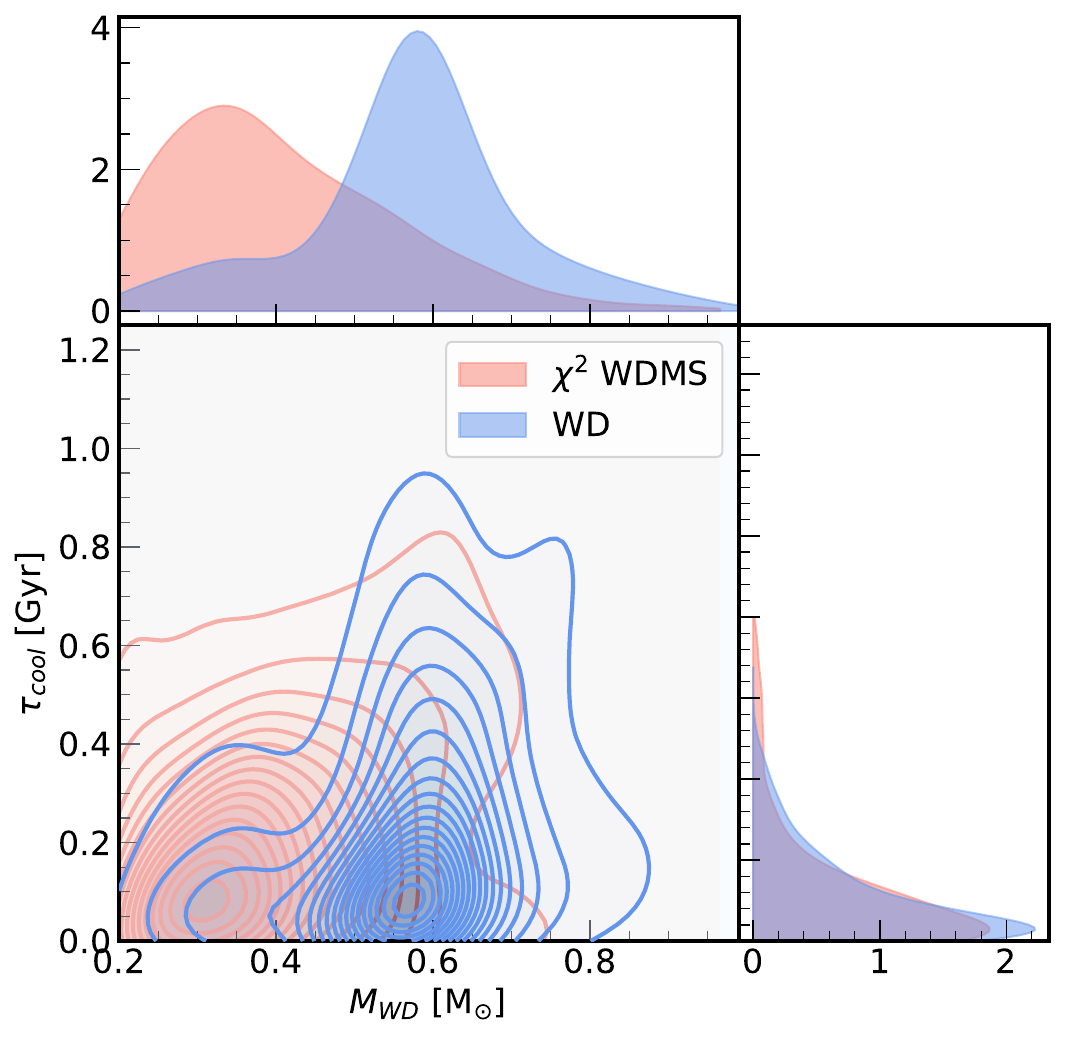}
     \includegraphics[width=0.49\linewidth]{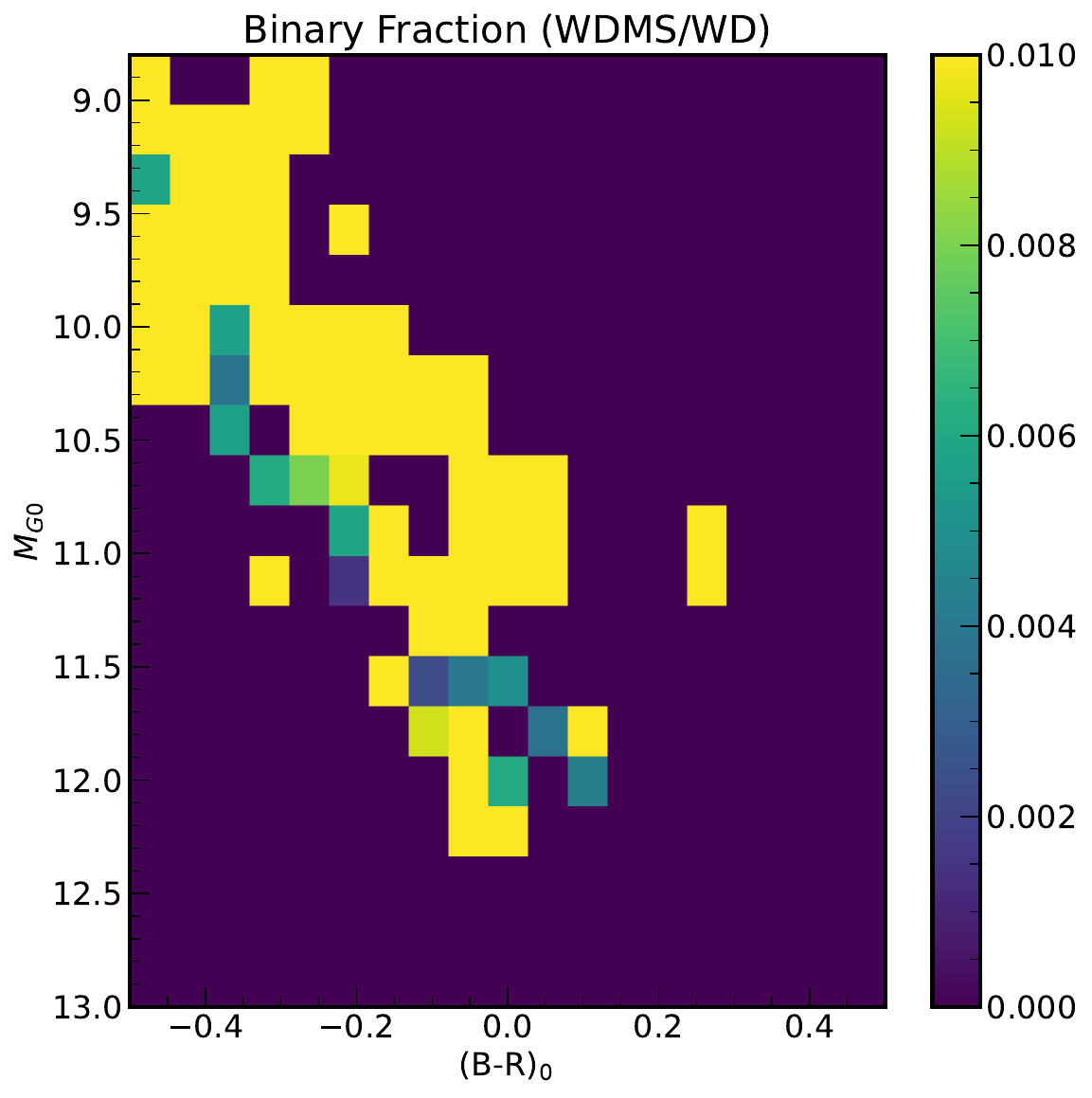}
     \includegraphics[width=0.49\linewidth]{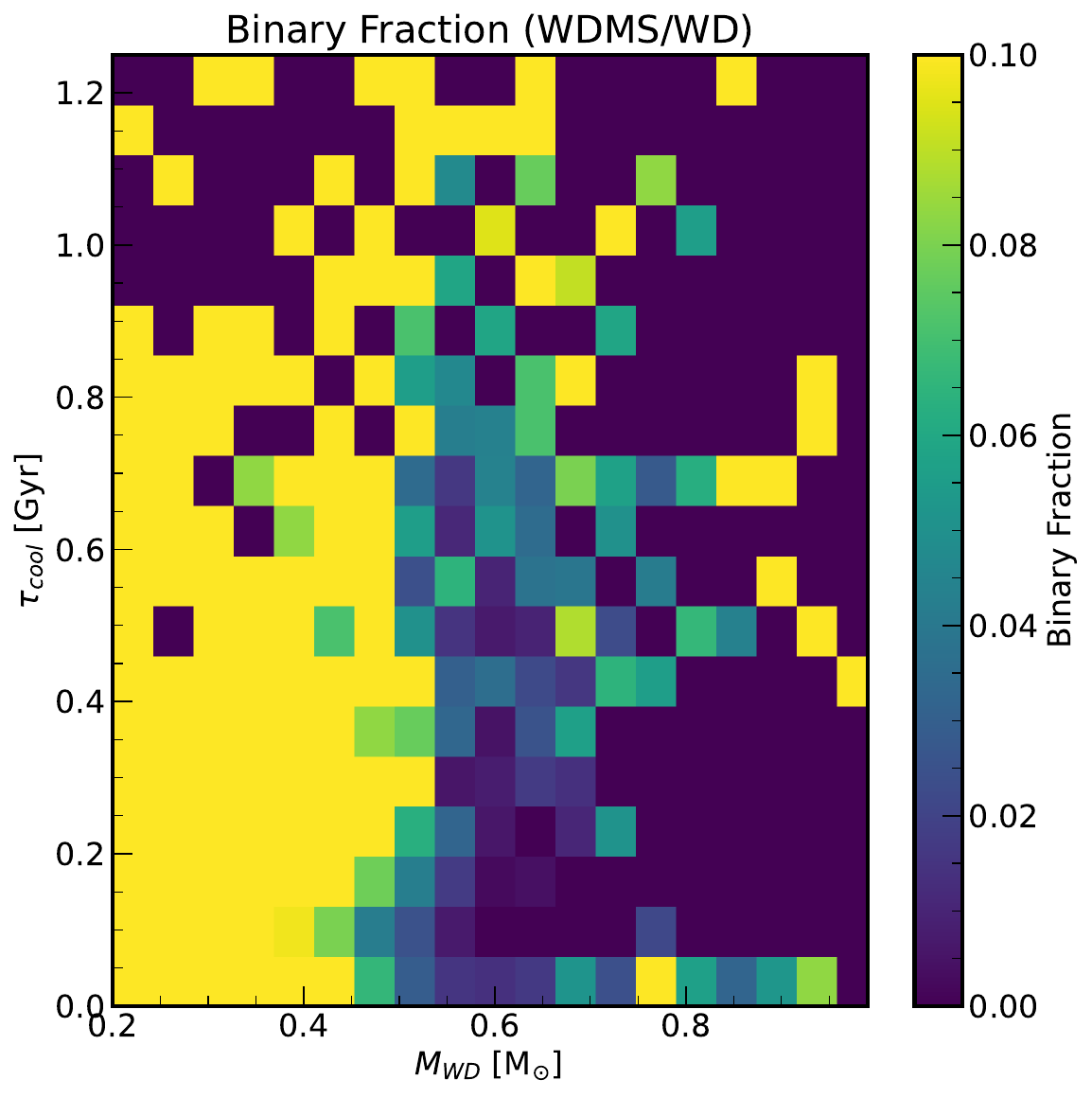}
    \caption{Comparison of properties between $\chi^2$-selected WDMS binaries (red) and single WDs (blue) from the \gaia\ sample \citep{Garcia-Zamora2023}. 
    Top-left panels show the CMD of $\chi^2$-selected WDMS and single WDs, contours indicating density distribution. 
    Top-right: The 2-d distributions between WD cooling age ($T_{\rm cool}$) and WD mass ($M_{\rm WD}$) of $\chi^2$-selected WDMS and single WDs. 
    Bottom panels show the binary fraction (WDMS/WD) as a function of CMD (left) and WD mass and cooling age (right), with the color scale indicating the fraction from 0 (purple) to 0.01 (yellow). 
    The $\chi^2$-selected WDMS sample shows a distinct excess of low-mass WDs compared to the field WD population, visible in both the mass distribution and binary fraction plots.}
    \label{fig:chi2_WD_stats}
\end{figure}

We determined the cooling ages and masses of the WD components in our $\chi^2$-selected WDMS binary sample using the \texttt{WD\_models}\footnote{\url{https://github.com/SihaoCheng/WD_models}} package, which implements the evolutionary models of white dwarfs from \cite{Bedard2020}. Figure~\ref{fig:chi2_WD_stats} shows the resulting distribution of masses and cooling ages in our \texttt{Chi2}-WDMS binary sample, in comparison with those from the ``field'' WD sample from \cite{Garcia-Zamora2023}.
For the field WD sample, we selected WDs with parallax $>$ 2 mas, $B$-band magnitude $<$ 18, and absolute magnitude $M_{G_0} < 13$. 
Both samples are distance-limited to approximately 500 pc (parallax $>$ 2 mas). 

At a given age, a WD's brightness is primarily determined by its cooling age rather than its mass, with the mass mainly affecting its temperature and color. 
As a result, our selection should primarily impact the system's age-mass combinations rather than the WD mass distribution itself. 

This overabundance of low-mass WDs in our WDMS sample is unlikely to be explained by selection effects for two key reasons:
1. Our selection function depends primarily on absolute G-band magnitude ($M_G$) rather than B-R color.
2. More massive WDs are typically hotter and therefore bluer, which should increase the color contrast with their MS companions. 
This enhanced contrast should make massive WDs more detectable in binary systems, contrary to our observed deficit.
The WD component in \cite{Rebassa2021}'s 100 pc WDMS sample peaks at $\sim 0.5$ \msun, which is also lower than the field WD mass peak, though the difference is not as pronounced as in our sample.

While our discovery of an excess of low-mass white dwarfs in WDMS binaries appears similar to the recent findings by \cite{Hallakoun2024}, who reported a deficit of massive white dwarfs in \gaia\ astrometric binaries.
Our $\chi^2$-selected sample focuses on spectroscopically fitting WDMS binaries through XP spectra, providing a different perspective from \cite{Hallakoun2024}'s study of astrometrically detected binaries with separations of $\sim$1 AU. 
\cite{Hallakoun2024} proposed two explanations for their observed deficit of massive white dwarfs:
First, the binary evolution channel, stable mass transfer during the asymptotic giant branch (AGB) phase of the WD progenitor, may preferentially produce lower-mass WDs due to the implicit mass ratio constraints required for stable mass transfer. 
They found that these systems show a mass-period relation that differs substantially from what is predicted for post-RGB interaction, suggesting a unique evolutionary pathway that could favor certain WD masses.

Second, the tight orbital separations ($\sim$1 AU) in their sample indicate these systems evolved through stable mass transfer, making it improbable for them to contain WDs that are merger products. 
This distinction is important since around 20\% of massive WDs with masses $\gtrsim 0.8\,M_\odot$ in the field likely formed through mergers \citep{Cheng2020}. 
The current orbital configuration at $\sim$1 AU would not have provided enough space for an inner binary to merge while maintaining the observed WDMS star separation, resulting in a lack of massive WD companions.

Our \texttt{Chi2}-WDMS sample consists of unresolved binaries whose separation distribution depends on distance, given \gaia's angular resolution limit of $\sim$2 arcsec. 
We assume a log-normal period distribution with $\log_{10} P = 5.03$ and $\sigma_{\log_{10} P} = 2.28$ (where $P$ is the orbital period in days) from \cite{raghavanSurveyStellarFamilies2010}. For our typical sample distance of 200 pc and assuming a total system mass of 0.8 \msun, approximately 15\% of binaries with $\log_{10} P > 6$ would be resolved by \gaia\ \citep{Liu2019}. 
This corresponds to separations of approximately 160 au, with the peak of the period distribution ($\log_{10} P = 5.03$) corresponding to $\sim$40 au. 

Therefore, our \texttt{Chi2} -WDMS binaries typically have wider separations than the astrometric binaries studied in \cite{Hallakoun2024}. 
The fact that we observe an excess of low-mass WDs even in these wider binaries lends support to the AGB stable mass transfer hypothesis. 
Since our sample includes systems wide enough to potentially host merger products, yet still shows a preference for low-mass WDs, this suggests that the binary evolution channel through AGB stable mass transfer plays a more dominant role in determining the WD mass distribution in these systems.

\section{Conclusion}
The \gaia\ DR3 BP/RP (XP) spectra provide a unique opportunity to identify and characterize WDMS binary systems, offering a middle ground between detailed but limited spectroscopic samples and cruder SED-based approaches. 
While spectroscopic surveys have been instrumental in discovering WDMS binaries, they are limited to a few thousand systems. 
Conversely, SED fitting alone often lacks the spectral resolution needed for reliable binary identification. 
The XP spectra's intermediate resolution necessitates new methodological approaches, as $\chi^2$ fitting alone proves insufficient for robust classification of these complex systems.

In this work, we have developed and validated a novel method for identifying white dwarf-main sequence (WDMS) binary systems using \gaia\ XP spectra. 
Our key contributions include:

We have constructed an innovative two-step methodology for WDMS binary identification. First, we developed neural network spectral emulators to model individual white dwarf and main sequence star spectra, then combined these to fit binary systems through direct $\chi^2$ minimization of the XP spectra. 
Second, recognizing the limitations of $\chi^2$ statistics alone for low-resolution spectra, we trained a Gaussian Process Classifier using both the $\chi^2$ statistics and synthetic colors derived from XP spectra as features.

This approach has yielded a comprehensive catalog of approximately 30,131 WDMS binary candidates. 
Our catalog provides two complementary samples optimized for different science goals: a $\chi^2$-selected sample of 1,649 WDMS binaries (\texttt{Chi2} sample) where both components contribute similarly to the optical flux, and a larger quality-filtered sample of 30,131 candidates identified through our GPC method (\texttt{FaintQC} sample). 
Cross-validation with \textit{GALEX} UV photometry confirms that $\sim$70\% of our GPC-selected candidates with $M_G > 7$ display clear UV excess, validating their binary nature even when they appear as single MS stars in optical bands. 
This represents an expansion of the \textit{all-sky} WDMS binaries over previous catalogs, which were discovered with spectroscopic surveys containing only a few thousand systems. 

Our analysis has also revealed interesting physical insights about WDMS binary evolution. 
In particular, we find an excess of low-mass white dwarfs (0.2--0.4 M$_{\odot}$) in WDMS binaries compared to the field white dwarf population. 
This suggests that binary evolution, particularly through stable mass transfer during the AGB phase, plays a crucial role in producing low-mass WDs.

The method developed here could be readily adapted to identify other types of binary systems in the \gaia\ database. 
This approach can be extended to low-resolution spectroscopic surveys such as SDSS-V and LAMOST, enabling efficient confirmation and characterization of candidates.
Our catalog provides a valuable resource for understanding stellar evolution in binary systems and will serve as a foundation for follow-up studies of individual systems of particular interest. 
Future spectroscopic surveys will be essential for confirming and characterizing these candidates, particularly systems where one component strongly dominates the optical flux.

\section{Acknowledgments}
JL thanks Andrew Saydjari, Kareem El-Badry, and Sahar Shahaf for helpful discussions.
We acknowledge support from the European Research Council through ERC Advanced Grant No. 101054731.
This work has made use of data from the European Space Agency (ESA) mission {\it Gaia} (\url{https://www.cosmos.esa.int/gaia}), processed by the {\it Gaia} Data Processing and Analysis Consortium (DPAC, \url{https://www.cosmos.esa.int/web/gaia/dpac/consortium}). Funding for the DPAC has been provided by national institutions, in particular the institutions participating in the {\it Gaia} Multilateral Agreement. YST is supported by the National Science Foundation under Grant AST-2406729.

\vspace{5mm}
\facilities{\textsl{Gaia}}


\software{
{\tt\string PyTorch} \citep{NEURIPS2019_9015},
{\tt\string Astropy} \citep{Astropy2018}, 
{\tt\string laspec} \citep{Zhang2021},
{\tt\string Scipy} \citep{2020SciPy-NMeth}, 
{\tt\string scikit-learn} \citep{scikit-learn}, 
TOPCAT \citep{2005ASPC..347...29T}, 
}

\clearpage 
\appendix
\section{Sampled XP spectra}\label{app:sample_xp}
To accurately process XP spectra, we must account for covariances between observed fluxes at different wavelengths. 
Our sampling range of 392--992~nm and step size of 10~nm were chosen to trim noisy spectral edges and ensure positive-definite covariance matrices \citep{zhang2023}.

The zero-point correction involves three components: the overall spectral zero-point and the relative zero-points for BP and RP spectra. We denote the covariance matrix in sample space as $\mathbf{C}_{\tilde{f}_{\text{gaia}}}$. 
Following \cite{zhang2023}, we inflate this matrix to avoid dominance by small uncertainties reported in \gaia\ DR3.

The final covariance matrix, $\mathbf{C}_{\tilde{f}_{\text{obs}}}$, is expressed as:
\begin{equation}
\mathbf{C}_{\vec{f}_{\text{obs}}} = \mathbf{C}_{\vec{f}_{\text{gaia}}} \! + \, \text{diag}\left(0.005 \vec{f}_{\text{gaia}}\right)^{\! 2} + 0.005^2 \vec{f}_{\text{gaia}} \vec{f}_{\text{gaia}}^T + 0.001^2 \vec{f}_{\text{BP}} \vec{f}_{\text{BP}}^T + 0.001^2 \vec{f}_{\text{RP}} \vec{f}_{\text{RP}}^T,
\end{equation}
where $\vec{f}_{\rm gaia}$ represents the sampled flux, and $\vec{f}_{BP}$ and $\vec{f}_{RP}$ are zero-padded vectors containing the sampled BP and RP spectra, respectively. The terms represent:
\begin{enumerate}
    \item Sample-space covariance matrix from \gaia\ DR3
    \item 0.5\% independent flux uncertainty at each wavelength
    \item 0.5\% uncertainty in the zero-point calibration of the overall spectral flux
    \item 0.1\% uncertainty in the zero-point calibration of the BP spectrum
    \item 0.1\% uncertainty in the zero-point calibration of the RP spectrum
\end{enumerate}

To ensure numerical stability, we diagonalize each covariance matrix using its eigendecomposition:
\begin{equation}
C_{\vec{f}{\rm obs}} = U D U^T,
\end{equation}
where $U$ is orthonormal and $D$ is diagonal. We impose a minimum value of $10^{-9}$ (in units of $10^{-36}\,\mathrm{W^2\,m^{-4}\,nm^{-2}}$) on the diagonal elements of $D$ (yielding $\hat{D}$) and define $L \equiv \hat{D}^{-1/2} U^T$. The $\chi^2$ statistic between predicted and observed fluxes is then:
\begin{equation}
\chi^2 = \Delta \vec{f}^T C_{\vec{f}_{\rm obs}}^{-1} \Delta \vec{f} = \left| L \Delta \vec{f} \right|^2,
\end{equation}
where $\Delta \vec{f} \equiv \vec{f}_{\rm pred} - \vec{f}_{\rm obs}$ represents the residuals between predicted and observed fluxes.

\section{Gaussian Process Classification}\label{app:GPC}

Gaussian Process Classification (GPC) is a probabilistic model that leverages Gaussian processes (GPs) to model a latent function, which is then transformed using a link function to predict class probabilities (e.g., in Section 6.4 of \citealt{Bishop2007}). 
The strength of GPs lies in their ability to capture complex, non-linear patterns while providing principled uncertainty estimates. 
For instance, in astronomical time-series analysis, GPs have proven invaluable in modeling stochastic signals like stellar variability or exoplanet transits \citep{Aigrain2023}, where flexibility and interpretability are crucial. 
Similarly, GPC brings these advantages to classification tasks, providing not only robust decision boundaries, but also measures of confidence that are vital in domains of noisy, high-dimensional data. 
By allowing for probabilistic boundaries, GPC accommodates uncertainties inherent in real-world data, making it an ideal choice for problems requiring both precision and adaptability.
A detailed online textbook of GPs amd GPC can be found at \url{https://computeastro.streamlit.app}.

GPC addresses several key challenges in machine learning: 
First, it provides flexible, non-linear decision boundaries without requiring manual feature engineering. 
Second, it offers probabilistic predictions that inherently capture uncertainty. 
Third, GPC automatically adapts model complexity to the density of the data.

Traditional approaches such as logistic regression enforce linear decision boundaries, necessitating careful feature engineering for non-linear relationships. 
On the other hand, NNs provide flexibility but often lack well-calibrated uncertainty estimates. 
GPC combines the non-linear adaptability of kernel methods with principled uncertainty quantification, bridging this gap.

In GPC, a latent function $a(x)$ is modeled as a Gaussian process and transformed through a sigmoid function $\sigma(\cdot)$ to obtain class probabilities. This two-stage approach captures complex, non-linear patterns while maintaining probabilistic interpretability and handling uncertainty in sparse data regions.

The formal model specification is:
\begin{equation}
    p(\mathbf{a}|\mathbf{X}) = \mathcal{N}(\mathbf{0}, \mathbf{K})  \quad \text{(GP Prior)},
\end{equation}
\begin{equation}
    p(t_n|x_n) = \sigma(a(x_n))^{t_n}(1-\sigma(a(x_n)))^{1-t_n} \quad \text{(Likelihood)},
\end{equation}
where \( \mathbf{K} \) is the kernel matrix that encodes similarities between data points, capturing their relationships under the Gaussian process prior. 
Here, \( t_n \in \{0,1\} \) represents the binary class label for the \( n \)-th data point, while \( x_n \) denotes the corresponding input features. 
The likelihood \( p(t_n | x_n) \) describes the probability of observing the label \( t_n \) for a given input \( x_n \), modeled using a Bernoulli distribution. 
The parameter of this distribution is the sigmoid function of the latent function \( a(x_n) \), given by \( \sigma(z) = \frac{1}{1+e^{-z}} \). 

The kernel function $k(x,x')$ determines the properties of the decision boundary. 
A common choice is the Radial Basis Function (RBF) kernel:
\begin{equation}
    k(x,x') = \exp\left(-\frac{\|x-x'\|^2}{2\ell^2}\right),
\end{equation}
where $\ell$ is the length scale parameter. 
This parameter controls the smoothness of the decision boundary, the influence range of training points, and the uncertainty level in data-sparse regions.

The classification likelihood in GPC is non-Gaussian, making the posterior analytically intractable. To address this, the Laplace approximation is used:

First, the maximum a posteriori (MAP) estimate $\mathbf{a}_{\text{MAP}}$ is obtained by maximizing:
    \begin{equation}
        \ln p(\mathbf{a}|\mathbf{X},\mathbf{t}) \propto -\frac{1}{2}\mathbf{a}^T\mathbf{K}^{-1}\mathbf{a} + \sum_{n=1}^N \ln p(t_n|a_n).
    \end{equation}
Second, the Hessian of the log-posterior is computed as:
    \begin{equation}
        \mathbf{H} = \mathbf{K}^{-1} + \text{diag}\{\sigma(a_n)(1-\sigma(a_n))\}
    \end{equation}
Third, the posterior is approximated as:
    \begin{equation}
        p(\mathbf{a}|\mathbf{X},\mathbf{t}) \approx \mathcal{N}(\mathbf{a}_{\text{MAP}}, \mathbf{H}^{-1})
    \end{equation}
    
The predictive distribution for a new point $x_*$ is:
\begin{equation}
    p(t_*=1|x_*, \mathbf{X}, \mathbf{t}) = \sigma\left(\frac{c}{\sqrt{1+\pi d^2/8}}\right),
\end{equation}
where:
\begin{equation}
    c = \mathbf{k}(x_*, \mathbf{X})(\mathbf{t}-\sigma(\mathbf{a}_{\text{MAP}})) \quad \text{(Mean prediction)},
\end{equation}
\begin{equation}
    d^2 = k(x_*, x_*) - \mathbf{k}(x_*, \mathbf{X})\left(\text{diag}\{\sigma(\mathbf{a}_{\text{MAP}})(1-\sigma(\mathbf{a}_{\text{MAP}}))\}^{-1} + \mathbf{K}\right)^{-1}\mathbf{k}(\mathbf{X}, x_*) \quad .
\end{equation}
The term $d^2$ increases with distance from training data, reducing confidence in sparse regions.
Furthermore, the model adapts based on data density, kernel parameters, and classification boundary difficulty.
Predictions naturally express uncertainty through probabilistic outputs and explicit variance estimates via $d^2$.

Effective GPC implementation requires attention to kernel selection, computational efficiency, and model validation. 
RBF kernels are well-suited for smooth decision boundaries, but composite kernels may capture more complex structures. 
Length scales and other hyperparameters can be optimized using maximum likelihood estimation.
The computational cost of GPC is $\mathcal{O}(N^3)$ for matrix operations, but sparse approximations can reduce this to $\mathcal{O}(M^2N)$, where $M \ll N$.

\begin{figure*}
    \centering
    \includegraphics[width=0.49\linewidth]{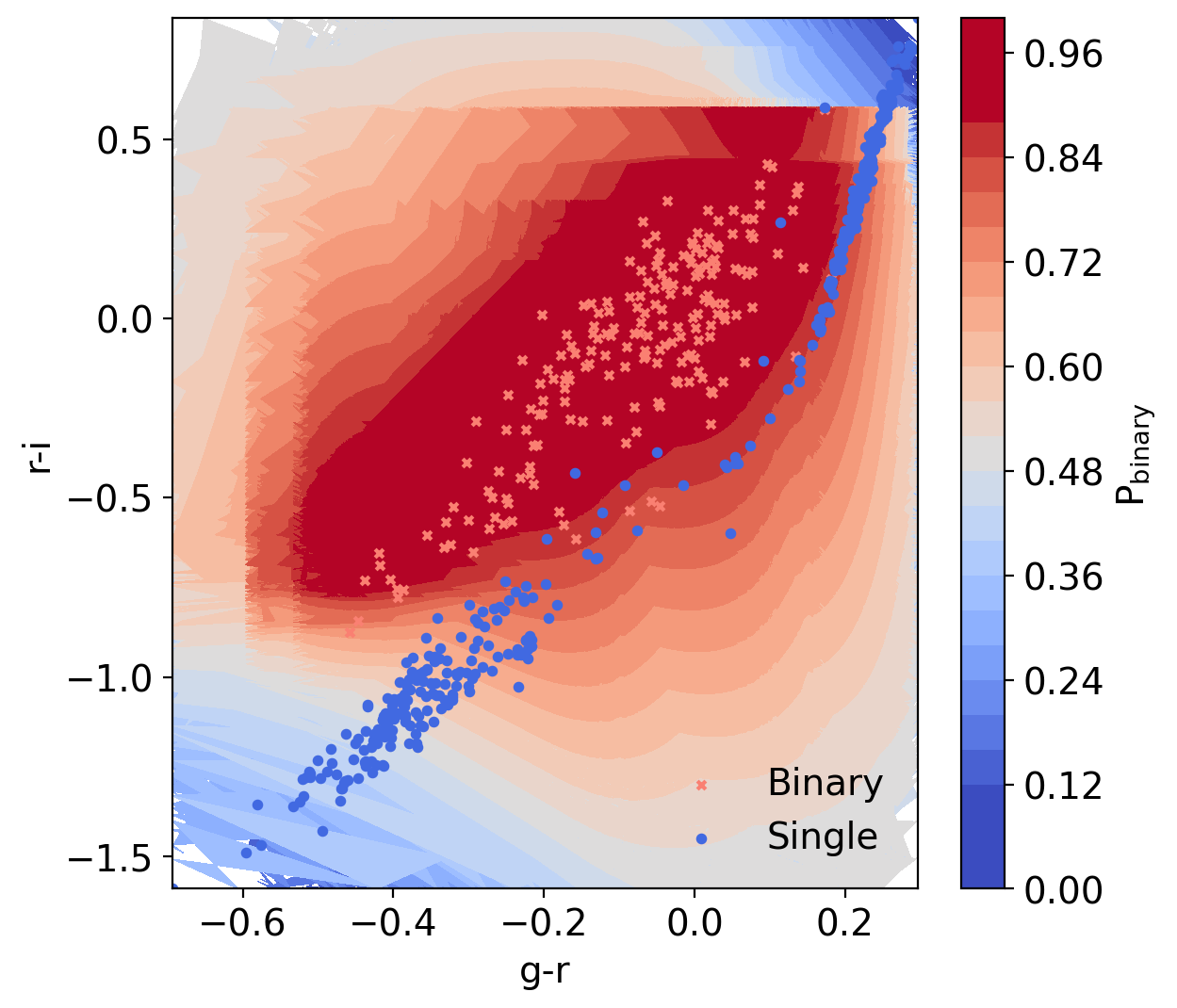}
    \includegraphics[width=0.49\linewidth]{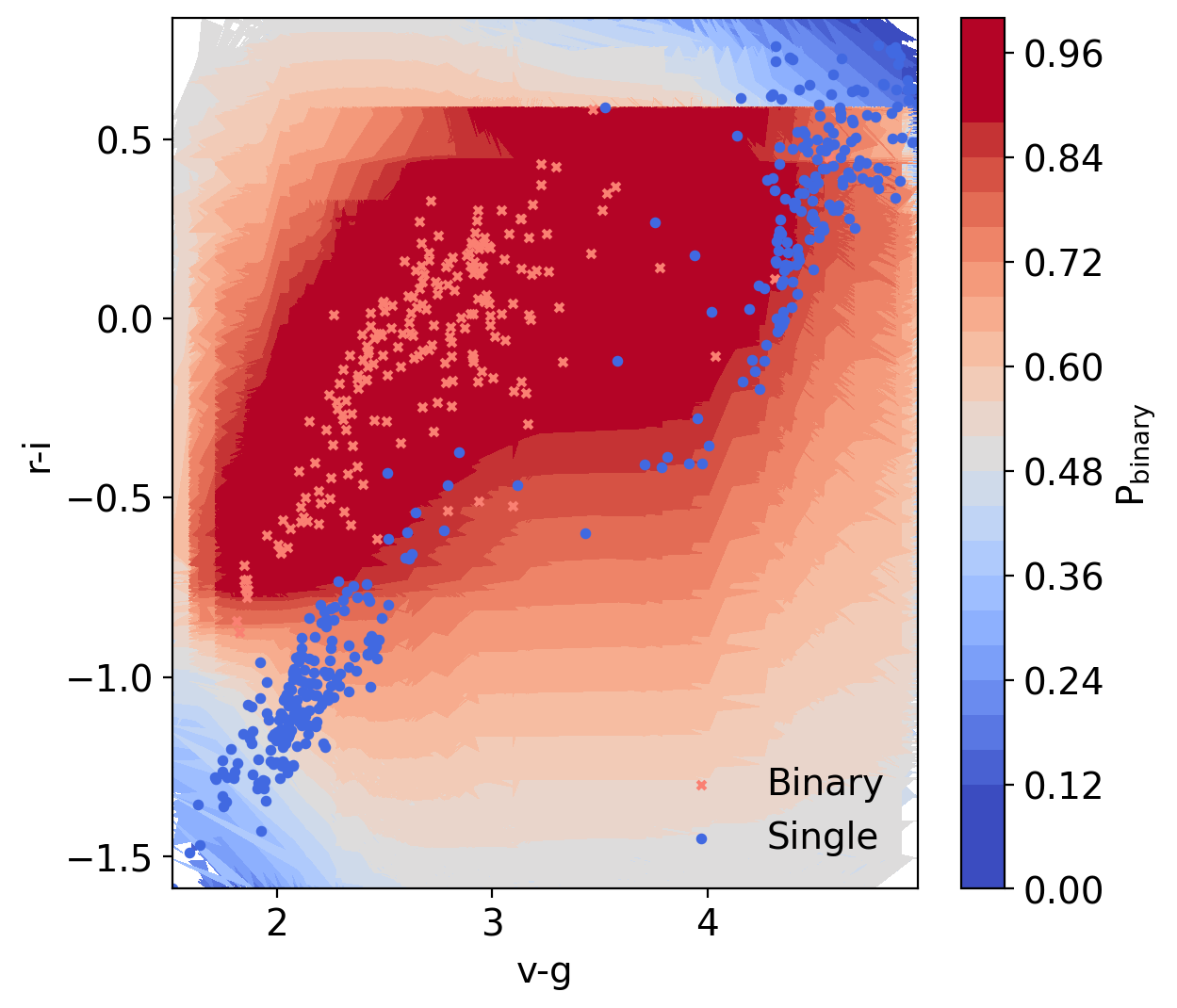}
    %
    \caption{
   Projections of the GPC predictions on single stars and WDMS binary systems using two example features: g-r vs r-i (top) and v-g vs r-i (bottom). 
    Blue points represent single stars while red crosses indicate binary systems. 
    The background contours show the predicted probability of being a binary system from the trained GPC model, ranging from 0 (blue) to 1 (red).}
    \label{fig:corner_mocktrain}
\end{figure*}

The training dataset comprised simulated single stars and binary systems, whose distributions were visualized in a multi-dimensional color-magnitude parameter space (Fig.~\ref{fig:corner_mocktrain}). 
The distinct separation between single and binary populations across multiple projections of this parameter space demonstrates the discriminative power of the selected photometric features. 
The GPC model learns the decision boundaries in this multi-dimensional space by fitting a Gaussian process to capture the non-linear relationships between the features.

To determine the optimal hyperparameters, we experiment with different RBF kernel length scales (0.1 and 1.0) and evaluate their performance using Receiver Operating Characteristic (ROC) curves. 
 Both length scales achieve good classification performance with Area Under the ROC Curve (AUC) values of 0.99 and 0.98.
While the difference is minimal, the length scale of 0.1 yields marginally better performance and is therefore selected for our final model.

\section{Normalizing Flow Sampling for Mock Binary Components}\label{app:nf}

To generate realistic training data for our GPC, we need to create mock WDMS binary systems that accurately represent the underlying stellar parameter distributions. 
This requires carefully modeling both the individual distributions of white dwarf (WD) and main sequence (MS) stars, as well as properly combining them to form binary systems. 
We accomplish this using Neural Spline Flows (NSF) \citep{durkan2019neural}.
Normalizing flows provide a framework for modeling complex probability distributions by transforming a simple base distribution, such as a standard normal, through a series of invertible mappings. They have already been applied successfully to tasks like modeling the Galactic potential \citep{Green2020, Green2023} and reconstructing the probability distribution of Milky Way disk stars in the space of multiple elemental abundances \citep{Ting2022}. 

Normalizing flows, particularly Neural Spline Flows (NSFs), excel at modeling stellar parameter distributions due to several key strengths. First, they can capture multi-modal and highly non-Gaussian distributions, which are commonly observed in stellar parameters. Second, they preserve exact likelihood computations, allowing for rigorous probabilistic modeling. Third, they leverage neural networks to learn highly flexible, non-linear transformations, which enable them to model complex structures in data efficiently. Finally, NSFs support efficient sampling of new data points from the learned distribution.

The "spline" component refers to the use of monotonic rational-quadratic splines as transformation functions. 
The splines provide substantially more flexible modeling capability compared to simpler alternatives like affine transformations, allowing the model to better capture complex parameter relationships.

Our approach models the joint distribution of three fundamental stellar parameters for each binary component. 
The color index $(B-R)_0$ provides crucial information about stellar temperature, while the absolute G magnitude $M_{G,0}$ serves as an indicator of overall luminosity. 
The absolute R magnitude $M_{R,0}$ helps constrain the stellar radius. We selected these parameters because they offer complementary information about the physical properties of stars while being directly observable quantities in our \gaia\ dataset.

The training dataset comprises two distinct samples: a clean sample of single WDs within 300 pc from Section \ref{sec:data}, and a sample of single MS stars within the same volume. 
We deliberately restrict training to stars within 300 pc to minimize Malmquist bias and other distance-dependent selection effects that could distort the learned distributions.


The NSF architecture consists of three primary transform layers, each incorporating a rational-quadratic spline with 8 bins. 
Each transform layer utilizes two hidden layers of 64 nodes with ReLU activation functions. 
A dedicated conditioning network handles stellar type information through one-hot encoded input (0 for WD, 1 for MS) and processes it through two hidden layers of 32 nodes. 
Training proceeds using the \texttt{Adam} with a learning rate of 0.001 and batch sizes of 256 samples. 
We implement early stopping with a patience of 10 epochs to prevent overfitting, while allowing training to continue for up to 100 epochs if needed. 

The NSF transforms a base distribution $p_Z(\mathbf{z})$ (standard normal) to the target distribution of stellar parameters $p_X(\mathbf{x}|c)$ through a bijective mapping $f_\theta$:

\begin{equation}
\mathbf{x} = f_\theta(\mathbf{z}; c), \quad \mathbf{z} \sim p_Z(\mathbf{z})
\end{equation}

where $\mathbf{x} = [(B-R)_0, M_{G,0}, M_{R,0}]$ and $c$ indicates the stellar type.

The conditional probability density is given by the change of variables formula:

\begin{equation}
p_X(\mathbf{x}|c) = p_Z(f^{-1}_\theta(\mathbf{x}; c)) \left|\det\frac{\partial f^{-1}_\theta(\mathbf{x}; c)}{\partial \mathbf{x}}\right|
\end{equation}

The model is trained by maximizing the log-likelihood:

\begin{equation}
\mathcal{L}(\theta) = \mathbb{E}_{(\mathbf{x}, c)\sim\mathcal{D}}[\log p_X(\mathbf{x}|c)]
\end{equation}
After training, we generate mock WDMS binary systems by first sampling the WD parameters from \( p_X(\mathbf{x}|c=0) \), and then sampling the MS parameters from \( p_X(\mathbf{x}|c=1) \). Finally, we combine the fluxes to create the unresolved binary properties.


\clearpage
\bibliography{main}{}
\bibliographystyle{aasjournal}


\end{document}